\documentclass{imsart}

\RequirePackage{amsthm,amsmath,amsfonts,amssymb}
\RequirePackage[authoryear]{natbib}
\RequirePackage[colorlinks,citecolor=blue,urlcolor=blue]{hyperref}
\RequirePackage{graphicx}
\RequirePackage[linesnumbered,ruled]{algorithm2e}
\RequirePackage{cases}
\RequirePackage{enumitem}
\RequirePackage{xcolor}
\RequirePackage{booktabs,array,tabularx,threeparttable, multirow, makecell, float}
\RequirePackage{subcaption}

\startlocaldefs
\newtheorem{theorem}{Theorem}

\newtheorem{proposition}[theorem]{Proposition}
\newtheorem{corollary}[theorem]{Corollary}
\usepackage[normalem]{ulem}

\theoremstyle{definition}

\newtheorem{remark}{Remark}

\DeclareMathOperator*{\argmin}{arg min}
\DeclareMathOperator*{\diag}{diag}

\DeclareMathOperator{\dist}{dist}

\SetKwFor{ParFor}{for}{do in parallel}{end}
\SetKwFunction{Sort}{sort}

\newcolumntype{Y}{>{\centering\arraybackslash}X}
\newcolumntype{P}[1]{>{\centering\arraybackslash}p{#1}}

\endlocaldefs

\begin{document}

\begin{frontmatter}
    \title{Scalable Ultra-High-Dimensional Quantile Regression with Genomic Applications}
    \runtitle{Scalable Ultra-High-Dimensional Quantile Regression}

    \begin{aug}
        \author[A]{\fnms{Hanqing}~\snm{Wu}\ead[label=e1]{hanqing.wu@stat.lu.se}},
        \author[B]{\fnms{Jonas}~\snm{Wallin}\ead[label=e2]{jonas.wallin@stat.lu.se}},
        \author[C]{\fnms{Iuliana}~\snm{Ionita-Laza}\ead[label=e3]{ii2135@cumc.columbia.edu}}
        \address[A]{Department of Statistics, Lund University, Lund, Sweden\printead[presep={,\ }]{e1}}
        \address[B]{Department of Statistics, Lund University, Lund, Sweden\printead[presep={,\ }]{e2}}
        \address[C]{Department of Biostatistics, Columbia University, New York, USA\printead[presep={,\ }]{e3}}
    \end{aug}

    \begin{abstract}
        Modern datasets arising from social media, genomics, and biomedical informatics are often heterogeneous and (ultra) high-dimensional, creating substantial challenges for conventional modeling techniques. Quantile regression (QR) not only offers a flexible way to capture heterogeneous effects across the conditional distribution of an outcome, but also naturally produces prediction intervals that help quantify uncertainty in future predictions. However, classical QR methods can face serious memory and computational constraints in large-scale settings. These limitations motivate the use of parallel computing to maintain tractability.
        While extensive work has examined sample-splitting strategies in settings where the number of observations $n$ greatly exceeds the number of features
        $p$, the equally important (ultra) high-dimensional regime
        $(p \gg n)$ has been comparatively underexplored. To address this gap, we introduce a feature-splitting proximal point algorithm, FS-QRPPA, for penalized QR in high-dimensional regime. Leveraging recent developments in variational analysis, we establish a Q-linear convergence rate for FS-QRPPA and demonstrate its superior scalability in large-scale genomic applications from the UK Biobank relative to existing methods. Moreover, FS‑QRPPA yields more accurate coefficient estimates and better coverage for prediction intervals than current approaches.  We provide a parallel implementation in the R
        package \texttt{fsQRPPA}, making penalized QR tractable on large-scale datasets.
    \end{abstract}

    \begin{keyword}
        \kwd{high-dimensional quantile regression}
        \kwd{feature splitting}
        \kwd{proximal point algorithm}
        \kwd{genomic applications}
    \end{keyword}

\end{frontmatter}


\section{Introduction}

Since the seminal work of \cite{koenkerRegressionQuantiles1978}, quantile regression (QR) has been a mainstream technique in statistics with applications across multiple fields, including epidemiology, economics, social sciences and genomics (\cite{yuQuantileRegressionApplications2003}, \cite{wangGenomewideDiscoveryBiomarkers2024}). QR extends classical linear regression (LR) from modeling the phenotypic mean to modeling the full phenotype distribution, allowing for the estimation of effects across the entire trait distribution. QR is a robust modeling technique, and one of its main advantages over LR is its ability to detect and characterize heterogeneity in effects across different quantiles of the outcome distribution. This quantile-specific modeling not only reveals patterns that classical LR cannot capture, but also provides a natural foundation for constructing prediction intervals by directly estimating the bounds of the outcome distribution. We are interested here in inference in the QR framework when data are (ultra) high-dimensional. For example, one relevant application we will focus on is the application to large-scale genomic datasets such as biobanks with tens (even hundreds) of thousands of individuals and millions of features (e.g. genetic variants). Scalable methods are critical in these settings because of the rapid expansion of biobanks worldwide and their  transformative impact on biomedical research.

\medskip

In these high-dimensional settings,  for both identification and feature selection considerations, we employ penalized QR (PQR). Let $y_i \in \mathbb R$ be a scalar response variable, and $\boldsymbol x_i = (1, x_{i1}, \dots, x_{ip})^\top \in \mathbb R^{p+1}$ be a $(p+1)$ dimensional feature vector including the intercept. We then solve the following optimization problem:
\begin{equation}
    \label{eq:vanilla_QR}
    \min_{\boldsymbol\beta} \quad \frac{1}{n}\sum_{i=1}^n \rho_\tau(y_i - \boldsymbol x_i^\top \boldsymbol\beta) + \Omega(\boldsymbol\beta).
\end{equation}
Here, $\boldsymbol \beta = (\beta_0, \beta_1, \dots, \beta_p)^\top \in \mathbb R^{p+1}$ is the coefficient vector, $\rho_\tau(z) = (\tau - 1\{z < 0\})z$ is the standard pinball loss for QR, and $\Omega(\boldsymbol \beta) $ is the penalty function. For example, $\Omega(\boldsymbol \beta)$ can be the weighted $\ell_1$ penalty, i.e. there exists some vector $\boldsymbol \lambda = (0, \lambda_1, \dots, \lambda_p)^\top \in \mathbb R^{p+1}$ of non-negative weights such that $\Omega(\boldsymbol \beta) = \sum_{i=1}^p \lambda_i|\beta_i|$.

\medskip

One major difficulty in solving \eqref{eq:vanilla_QR} lies in the non-differentiability of the pinball loss. There are various methods for convex penalized QR, including interior point methods, coordinate descent, gradient descent based on smoothing, and alternating direction method of multipliers (ADMM). With ADMM, instead of directly minimizing (\ref{eq:vanilla_QR}), one decouples the pinball loss from the penalty, decomposing the original difficult minimization problem into easier subproblems. In the context of high-dimensional QR, ADMM's inherent suitability for parallelization makes it particularly competitive in terms of speed, estimation accuracy, and feature selection relative to other methods (e.g. \cite{boydDistributedOptimizationStatistical2010}, \cite{yuADMMPenalizedQuantile2017}, \cite{guADMMHighDimensionalSparse2018}).

\medskip

Nowadays, researchers are often faced with a (ultra) high-dimensional design matrix $\boldsymbol X \in \mathbb R^{n\times ({p+1})}$, where either $n$ or $p$, or both are excessively large, i.e. on the order of hundreds of thousands, or even millions. This leads to memory and computational bottlenecks. Via ADMM it is possible to further split the matrix $\boldsymbol X$ into multiple tractable partitions, distribute these partitions and the corresponding subsets of variables into multiple cores or local machines and update the variables simultaneously (see Chapter 8 in  \cite{boydDistributedOptimizationStatistical2010} for several ADMM applications to high-dimensional regression problems). The idea of parallelization via ADMM has also been extended to QR   (\cite{yuADMMPenalizedQuantile2017}). Indeed, there is a substantial and growing literature on how the distributed version of ADMM can be deployed to solve QR problems where $n$ is excessively large through splitting $\boldsymbol X$ row-wise or, in other words, sample-splitting  (see, e.g., \cite{yuADMMPenalizedQuantile2017}, \cite{fanPenalizedQuantileRegression2021}, \cite{liuExtendedADMMGeneral2024} \cite{wuPartitionInsensitiveParallelADMM2024}, \cite{wuParallelADMMAlgorithm2025}, \cite{mirzaeifardDecentralizedSmoothingADMM2025}).

\medskip

Much less research has addressed feature splitting, i.e. partitioning $\boldsymbol X$ along its columns in settings where the number of features $p$ is extremely large, even though this situation is common in many scientific fields, particularly in genomics where high-throughput assays routinely measure millions of features at a time. For example, the UK Biobank (\cite{bycroftUKBiobankResource2018}) is a large-scale population resource with genotype and phenotype data on roughly half a million individuals  with millions of measured genetic variables. One potential reason is that in sample-splitting, each parallel worker solves exactly the same subproblem as in the centralized case, and one only needs to find a suitable way to aggregate quantities such as gradients, local estimators, or residuals. By contrast, feature splitting is ``harder" in the sense that we need to handle strong coupling across blocks and maintain global residual information.  \cite{yuADMMPenalizedQuantile2017} briefly discussed a direct extension of the canonical two-block ADMM scheme with feature splitting. However, the lack of theoretical justifications and convergence guarantees renders their approach essentially heuristic. Moreover, their numerical studies considered only cases with $p=1000$, which is far below the (ultra) high-dimensional scales encountered in real-world applications. In light of this, and also to avoid the pitfall that the direct extension of canonical two-block ADMM does not necessarily converge (\cite{chenDirectExtensionADMM2016}), \cite{wenFeaturesplittingAlgorithmsUltrahigh2025} proposed a three-block ADMM algorithm (FS-QRADMM) for feature-splitting PQR with convergence guarantee and numerical efficiency. However, as pointed out in \cite{wuPartitionInsensitiveParallelADMM2024}, FS-QRADMM introduces a total number of $2Gn$ auxiliary variables, and updates $p + 1 + (3G-1)n$ variables per iteration, where $G$ is the number of partitions. In addition, the iterative cycle of FS-QRADMM entails multiple global synchronizations and communication among processors or local machines, which can be expensive at scale. These factors can make the computational effectiveness of FS-QRADMM less pronounced when $G$ or $n$ is large. Furthermore, the linear convergence rate in \cite{wenFeaturesplittingAlgorithmsUltrahigh2025} is expressed in terms of a relatively complicated metric combining iterate errors with successive differences, making the convergence behavior of the algorithm less directly interpretable.

\medskip

Last but not least, we note a crucial gap between theoretical methodological advancements and available computational implementations. To the best of our knowledge, a parallelized implementation for feature-splitting PQR is currently unavailable. Furthermore, efficient parallel implementations are scarce even for the widely studied sample-splitting algorithms; existing packages often rely on sequential loops to simulate parallel updates rather than exploiting true multi-core or distributed computing. This lack of efficient implementation hinders the practical application of these splitting algorithms to (ultra) high-dimensional data.

\medskip

\paragraph*{Motivating application: biobank scale genomics data.}
Our main motivation comes from applications to large-scale genomic studies. For the past twenty years, genome-wide association studies (GWAS) have been the dominating approach to identify genotype-phenotype associations and for genomic trait prediction. GWAS is a large-scale, hypothesis-free study that tests for associations between millions of genetic variants measured across the entire genome and a phenotype of interest. Traditionally, GWAS were conducted through consortia-based meta-analyses, combining multiple conventional cohorts to achieve large sample sizes. Today, large population-scale biobanks, such as
the UK Biobank, the All of Us Research Program in the United States
and other global biobanks across the world, provide unified, deeply phenotyped datasets with tens or hundreds of thousands of participants, enabling GWAS on an unprecedented scale within a single resource. The analysis of biobank-scale datasets poses major computational and memory challenges due to massive sample sizes and high dimensionality.

\medskip

Essentially almost all GWAS studies are based on linear regression models (\cite{uffelmannGenomewideAssociationStudies2021}). These models adopt a static view of genetic effects, implicitly assuming that genetic influences remain constant across environments. Such an assumption overlooks well-documented complexities in genotype–phenotype relationships, including gene–gene (G×G) and gene–environment (G×E) interactions, where “environment’’ is broadly construed to encompass a wide spectrum of biological and external factors (\cite{pazokitoroudiScalableRobustVariance2024}, \cite{mackayEpistasisQuantitativeTraits2014}, \cite{mackayPleiotropyEpistasisGenetic2024}). Such underlying interactions induce heterogeneity in genetic associations and linear models in GWAS are not well-suited to decipher such heterogeneous associations (\cite{wangGenomewideDiscoveryBiomarkers2024}). While non-linear models, including deep learning approaches, have been proposed for GWAS, their large-scale adoption has been hindered by key challenges such as high computational demands, instability in variable selection, lack of robust inferential tools and limited interpretability. In contrast, statistical methods like QR offer a principled way to move beyond linear assumptions, enabling the detection of heterogeneous and context-dependent genetic associations. QR retains many of the core advantages of linear regression, which is central to GWAS: it is a regression-based framework that supports covariate adjustment and accounts for sample relatedness, yields interpretable coefficient estimates at specific quantile levels, and provides strong statistical guarantees, features that are often lacking in deep learning approaches.

\medskip

Beyond genetic discovery, genomic trait prediction is a key focus in precision medicine, as well as in animal and plant breeding, with the goal to develop accurate phenotype prediction models that leverage an individual’s genome-wide genetic profile, commonly summarized as a polygenic risk score. QR offers a natural way to quantify uncertainty around individual predictions by providing prediction intervals that give the lower and upper bounds where the response lies with high probability.
\medskip

Despite their attractive properties, QR methods have seen limited use in genetics, mostly restricted to small-scale studies, leaving substantial potential untapped. A major barrier is the absence of scalable QR approaches capable of managing the core challenges of large-scale GWAS data, including (ultra) high-dimensionality with hundreds of thousands of samples and tens of millions of variants. These limitations motivate the methodological developments introduced in this manuscript.

\medskip

\paragraph*{Our contributions. }
Inspired by recent work of \cite{wuFeatureSplittingParallel2025} on a feature-splitting proximal point algorithm (PPA) for Dantzig selectors, in this work  we propose FS-QRPPA, a novel algorithm designed for PQR with weighted $\ell_1$ regularization that also naturally accommodates non-convex penalties such as the Smoothly Clipped Absolute Deviation (SCAD) (\cite{fanCommentsWaveletsStatistics1997}) and the Minimax Concave Penalty (MCP) (\cite{zhangNearlyUnbiasedVariable2010}). Structurally similar to the canonical two-block ADMM, FS-QRPPA uses only $2n$ auxiliary variables, offering benefits in terms of both memory usage and computational cost. Moreover, each iteration of FS-QRPPA consists of simpler sequential updates than FS-QRADMM. This, in turn, substantially decreases the synchronization and communication overhead. It is worth noting that \cite{wuFeatureSplittingParallel2025} report only a worst-case sublinear convergence rate of PPA for Dantzig selectors. By contrast, building upon recent theoretical development of variable-metric PPA, we establish a Q-linear convergence rate for our algorithm in a simple form, which provides clear insight into the algorithm's behavior. From a practical standpoint, another contribution of this work is the development of a multi-core parallel R implementation of FS-QRPPA, provided as the open-source R package \texttt{fsQRPPA}, developed using \texttt{RcppParallel} and \texttt{RcppArmadillo}. Although several distributed QR algorithms have been proposed, efficient parallel implementations remain scarce, limiting their use in large-scale modern applications. As a result, most existing applications have focused on datasets of relatively modest size. To address this limitation, we demonstrate the scalability of FS-QRPPA by applying it to genome-wide data from the UK Biobank.

\medskip

\paragraph*{Organization.} The remainder of this article is organized as follows. In Section \ref{sec:lagrangian} we formulate the general PQR problem with a focus on the weighted $\ell_1$ penalty case and derive its corresponding Lagrangian. In Section \ref{sec:ADMM} we review existing ADMM solutions and discuss their computational shortcomings. In Section \ref{sec:FS-QRPPA} we detail the proposed FS-QRPPA framework, the convergence properties of which are formally established in Section \ref{sec:Q-linear}. In Section \ref{sec:non-convex} we describe how FS-QRPPA can be seamlessly extended to handle PQR with non-convex penalties. We validate the scalability and utility of our approach through extensive numerical studies and a biobank-scale application in Section \ref{sec:numerical_studies}. Finally, in  Section \ref{sec:discussion} we conclude with a discussion on future directions. All technical proofs and additional details are provided in the Appendix.

\medskip

\paragraph*{Notations and definitions.}
We summarize here the notations and definitions used throughout the paper. We denote by $q_{Y\mid X}(\tau \mid \boldsymbol x) = \inf\{y \in \mathbb R : F_{Y\mid X}(y \mid \boldsymbol x) \ge \tau\}$ the $\tau$-th conditional quantile of $Y$ given $X = \boldsymbol x$, and, when no confusion can arise, we write $q_\tau(\boldsymbol x)$ for short. Let $\mathbb{R}^m$ denote the $m$-dimensional Euclidean space, and let $1\{\cdot\}$ denote the indicator function, which equals $1$ if the condition inside holds and $0$ otherwise. The symbol $\boldsymbol{0}$ represents a zero vector or matrix of appropriate dimension. Similarly, we let $\boldsymbol 1$ represent a vector of all ones of suitable dimension. For a scalar $v \in \mathbb{R}$, we write $v_+ = \max\{v,0\}$, and for a vector $\boldsymbol{v} = (v_1,\dots,v_m)^\top \in \mathbb{R}^m$, we define $\boldsymbol{v}_+ = \big((v_1)_+,\dots,(v_m)_+\big)^\top$. Given $\boldsymbol{v}_1, \boldsymbol{v}_2 \in \mathbb{R}^m$, their element-wise (Hadamard) product is denoted by $\boldsymbol{v}_1 \circ \boldsymbol{v}_2$. For a real symmetric matrix $\boldsymbol{M} \in \mathbb{R}^{m \times m}$, we denote by $\Lambda_{\max}(\boldsymbol{M})$ and $\Lambda_{\min}(\boldsymbol{M})$ its largest and smallest eigenvalue, respectively, and if $\boldsymbol{M}$ is positive definite we write $\boldsymbol{M} \succ 0$. The notation $\diag(d_1,\dots,d_m)$ refers to the diagonal (or block-diagonal) matrix with diagonal entries (or blocks) $d_1,\dots,d_m$, while $\boldsymbol{I}_m$ denotes the $m \times m$ identity matrix. We use $\|\cdot\|_1$ and $\|\cdot\|_2$ to denote the $\ell_1$ and $\ell_2$ norms on $\mathbb{R}^m$, respectively. Let $\boldsymbol B \in \mathbb{R}^{m\times m}$ be a symmetric positive definite matrix. For any $\boldsymbol{v} \in \mathbb{R}^m$ and $\mathcal C \subset \mathbb{R}^m$, define the $\boldsymbol B$-induced norm $\|\boldsymbol{v}\|_{\boldsymbol B} = \sqrt{\boldsymbol{v}^\top \boldsymbol B \boldsymbol{v}}$ and the corresponding distance $\dist_{\boldsymbol B}(\boldsymbol{v}, \mathcal C) = \inf_{\boldsymbol{s} \in \mathcal C} \|\boldsymbol{v} - \boldsymbol{s}\|_{\boldsymbol B}$. In particular, when $\boldsymbol B = \boldsymbol I_m$ is the identity matrix, $\dist_{\boldsymbol B}(\boldsymbol{v}, \mathcal C) = \dist(\boldsymbol{v}, \mathcal C) = \inf_{\boldsymbol{s} \in \mathcal C} \|\boldsymbol{v} - \boldsymbol{s}\|_2$ is the Euclidean distance. Finally, let $f: \mathbb R^m \to \overline{\mathbb R}$. For $\boldsymbol v\in\mathbb R^m$, we denote by $\partial f(\boldsymbol v)$ the subdifferential of $f$ at $\boldsymbol v$, whenever it is well defined.

\section[Penalized Quantile Regression and its Lagrangian]{Penalized Quantile Regression and its Lagrangian}
\label{sec:lagrangian}

In general, the PQR problem \eqref{eq:vanilla_QR} can be equivalently reformulated as the following constrained optimization problem:
\begin{equation}
    \label{eq:reformulated_QR}
    \begin{aligned}
        \min_{{\boldsymbol\beta}, {\boldsymbol z}} \quad & \hat Q_{\tau}(\boldsymbol z) + \Omega(\boldsymbol\beta),       \\
        \text{s.t.}                                \quad & \boldsymbol z = \boldsymbol y - \boldsymbol X\boldsymbol\beta,
    \end{aligned}
\end{equation}
where $\hat Q_\tau(\boldsymbol z) = \frac{1}{n}\sum_{i=1}^n \rho_\tau(z_i)$.

\medskip

The Lagrangian for \eqref{eq:reformulated_QR} is given by:
\begin{equation}
    \label{eq:main_lagrangian}
    \mathcal L(\boldsymbol \beta, \boldsymbol z; \boldsymbol \theta)= \Omega(\boldsymbol\beta) + \hat Q_\tau(\boldsymbol z) + \boldsymbol \theta^\top\left(\boldsymbol y - \boldsymbol X\boldsymbol\beta - \boldsymbol z\right),
\end{equation}
where $\boldsymbol \theta \in \mathbb R^n$ is the dual vector. When $\Omega(\boldsymbol\beta)$ is a convex penalty, the original QR problem \eqref{eq:vanilla_QR} can be equivalently viewed as finding the saddle point of Lagrangian \eqref{eq:main_lagrangian}. This is a standard result (see, e.g., \cite{rockafellarVariationalAnalysis1998}), which we state in the following proposition for completeness. More details are provided in Appendix C.
\begin{proposition}
    \label{prop:min_saddle_points}
    Assume $\Omega(\boldsymbol \beta)$ is convex. Then $\bar{\boldsymbol u} = (\bar{\boldsymbol \beta}^\top, \bar{\boldsymbol z}^\top)^\top \in \mathbb R^{p+n}$ solves the minimization problem \eqref{eq:reformulated_QR} if and only if there exists some $\bar{\boldsymbol \theta} \in \mathbb R^n$ such that $(\bar{\boldsymbol u}, \bar{\boldsymbol \theta})$ is the saddle point of Lagrangian $\mathcal L$ in \eqref{eq:main_lagrangian}, i.e., for all $\boldsymbol u \in \mathbb R^{p+n}, \boldsymbol\theta \in \mathbb R^{n}$, we have
    \begin{equation*}
        \mathcal L(\bar{\boldsymbol u}; \boldsymbol \theta) \le \mathcal L(\bar{\boldsymbol u}; \bar{\boldsymbol \theta}) \le \mathcal L(\boldsymbol u; \bar{\boldsymbol \theta}).
    \end{equation*}
\end{proposition}

\medskip

In the following two sections, we focus on the specific case where $\Omega(\boldsymbol\beta)$ is the weighted $\ell_1$ penalty, with the LASSO penalty being its special case.
In particular,
\begin{equation}
    \label{eq:weight_l1_penalty}
    \Omega(\boldsymbol \beta) = \|\boldsymbol \lambda \circ \boldsymbol\beta\|_1 = \sum_{j=1}^p \lambda_j |\beta_j|,
\end{equation}
where $\lambda_0 = 0$ and $\lambda_j \ge 0$ for $j = 1, \dots, p$.   There is a rich literature on PQR with penalty \eqref{eq:weight_l1_penalty}. This is not only due to its direct connection to the popular LASSO penalty, but also due to the fact that, once the weighted $\ell_1$ PQR can be solved efficiently, high-quality solutions to PQR with general folded-concave penalty can be obtained via local linear approximation (LLA) (\cite{zouOnestepSparseEstimates2008}); we discuss such nonconvex penalties later in the manuscript.

\section{ADMM for PQR}
\label{sec:ADMM}
\subsection{The canonical ADMM and its multi-block extension for PQR}
Directly finding the saddle point of \eqref{eq:main_lagrangian} can be a difficult problem. Standard two-block ADMM attempts to find the saddle point of the original Lagrangian \eqref{eq:main_lagrangian} by considering its augmented version
\begin{equation}
    \label{eq:augmented_lagrangian}
    \mathcal L_\rho(\boldsymbol \beta, \boldsymbol z; \boldsymbol \theta)= \Omega(\boldsymbol\beta) + \hat Q_\tau(\boldsymbol z) + \boldsymbol \theta^\top\left(\boldsymbol y - \boldsymbol X\boldsymbol\beta - \boldsymbol z\right) + \frac{\rho}{2}\|\boldsymbol y - \boldsymbol X\boldsymbol\beta - \boldsymbol z\|^2_2,
\end{equation}
where $\rho > 0$ is augmented Lagrangian parameter. A saddle point of \eqref{eq:augmented_lagrangian}, which is also the saddle point of \eqref{eq:main_lagrangian}, can be obtained iteratively. Specifically, let $\boldsymbol \beta^k, \boldsymbol z^k, \boldsymbol \theta^k$ be the updates obtained at $k$-th iteration. Then the standard ADMM algorithm looks as follows:
\begin{equation}
    \label{eq:admm_update}
    \begin{cases}
        \boldsymbol\beta^{k+1}  & = \argmin_{\boldsymbol \beta}  \mathcal L_\rho(\boldsymbol \beta, \boldsymbol z^k; \boldsymbol \theta^k)=\argmin_{\boldsymbol \beta} \Omega(\boldsymbol\beta) - \boldsymbol\theta^k{}^\top\boldsymbol X\boldsymbol\beta + \frac{\rho}{2}\|\boldsymbol y - \boldsymbol X\boldsymbol\beta - \boldsymbol z^k\|^2_2 \\
        \boldsymbol z^{k+1}     & = \argmin_{\boldsymbol z}  \mathcal L_\rho(\boldsymbol \beta^{k+1}, \boldsymbol z; \boldsymbol \theta^k)=\argmin_{\boldsymbol z}   \hat Q_\tau(\boldsymbol z) - \boldsymbol\theta^k{}^\top\boldsymbol z + \frac{\rho}{2}\|\boldsymbol y - \boldsymbol X\boldsymbol\beta^{k+1} - \boldsymbol z\|^2_2             \\
        \boldsymbol\theta^{k+1} & = \boldsymbol \theta^k + \rho(\boldsymbol y - \boldsymbol X\boldsymbol\beta^{k+1} - \boldsymbol z^{k+1})
    \end{cases}
\end{equation}
For a general design matrix $\boldsymbol X$, the update for $\boldsymbol \beta^{k+1}$ lacks a closed-form solution. It can be solved through coordinate descent (\cite{yuADMMPenalizedQuantile2017}, \cite{guADMMHighDimensionalSparse2018}), by introducing more auxiliary variables (\cite{yuParallelAlgorithmLargeScale2017}), or by adding a linearization term (\cite{guADMMHighDimensionalSparse2018}).
The memory and computational bottleneck becomes particularly severe when $\boldsymbol X$ is prohibitively large, and in extreme cases, the full matrix $\boldsymbol X$ cannot be loaded into memory on a single machine. Under these circumstances, the repetitive direct evaluation of $\boldsymbol X\boldsymbol\beta$ ($O(np)$ operations) required for iterative optimization becomes computationally impractical. Parallel computing makes it possible to partition the intractable huge problem into  feasible sub-tasks. When the number of features $p$ is large, a feature-splitting for the update of $\boldsymbol \beta^{k+1}$ is desirable. In our setting of weighted $\ell_1$ penalty, a splitting is possible by introducing more primal auxiliary variables, as $\Omega(\boldsymbol \beta)$ can be viewed as the sum of penalties imposed on each coordinate and hence ``separable". However, such an extension is not straightforward as the resulting multi-block ADMM may not necessarily converge (\cite{chenDirectExtensionADMM2016}). To this end, \cite{wenFeaturesplittingAlgorithmsUltrahigh2025} proposed a feature-splitting ADMM for PQR (FS-QRADMM) where $\Omega(\boldsymbol \beta)$ is the weighted $\ell_1$-penalty utilizing three-block ADMM with convergence guarantee\footnote{We only cover FS-QRADMM-prox in \cite{wenFeaturesplittingAlgorithmsUltrahigh2025} where a variant without the proximal term is also introduced, i.e., FS-QRADMM-CD. However, FS-QRADMM-CD does not have explicit-form update for $\boldsymbol \beta$ and relies on coordinate descent, which could be less performant compared with FS-QRADMM-prox. More importantly, FS-QRADMM-CD is lacking in convergence guarantee.}. Note that the design matrix $\boldsymbol X$ can be partitioned column-wise into $G$ parts such that
\begin{equation}
    \label{eq:decompose_x}
    \boldsymbol X = (\boldsymbol X_1, \dots, \boldsymbol X_G),  \text{ where } \boldsymbol X_g \in \mathbb R^{n\times p_g},\quad g = 1, \dots, G \text { and } \sum_{g=1}^G p_g = p+1.
\end{equation}
Coefficient vector $\boldsymbol \beta$ can be partitioned correspondingly, i.e.
\begin{equation*}
    \boldsymbol \beta = \left(\boldsymbol \beta_1^\top, \dots, \boldsymbol \beta_G^\top\right)^\top, \text{ where } \boldsymbol\beta_g \in \mathbb R^{p_g}.
\end{equation*}
We have that $\boldsymbol X \boldsymbol\beta = \sum_{g=1}^G \boldsymbol X_g\boldsymbol \beta_g$ and
\begin{equation}
    \label{eq:decompose_omega}
    \Omega(\boldsymbol\beta) = \sum_{g=1}^G\Omega_g(\boldsymbol\beta_g), \text{ where } \Omega_g(\boldsymbol \beta_g) =\|\boldsymbol \lambda_g \circ \boldsymbol \beta_g\|_1= \sum_{j=1}^{p_g} \lambda_{j, g} |\beta_{j, g}|.
\end{equation}

In the standard two-block ADMM formulation, the quadratic term containing $\boldsymbol{X}\boldsymbol{\beta}$ couples different $\boldsymbol \beta_g$, which prevents parallelization. \cite{wenFeaturesplittingAlgorithmsUltrahigh2025} introduce additional slack variables $\boldsymbol w_g \in \mathbb R^n,\ g=2, \dots, G$ to achieve decoupling. They reformulate the original PQR problem \eqref{eq:vanilla_QR} as
\begin{equation}
    \label{eq:reformulated_QR_fsadmm}
    \begin{aligned}
        \min_{{\boldsymbol\beta}, {\boldsymbol z}, {\boldsymbol{\omega}_g}} \quad & \hat Q_{\tau}(\boldsymbol z) + \Omega(\boldsymbol\beta),                                                \\
        \text{s.t.}                                \quad                          & \boldsymbol z + \boldsymbol X_1\boldsymbol\beta_1 + \sum_{g=2}^G \boldsymbol{\omega}_g= \boldsymbol y , \\
                                                                                  & \boldsymbol X_g \boldsymbol \beta_g = \boldsymbol{\omega}_g, \quad g=2, \dots, G.
    \end{aligned}
\end{equation}
The corresponding augmented Lagrangian reads
\begin{equation}
    \label{eq:fsadmm_lagrangian}
    \begin{aligned}
        \mathcal{L}_\phi(\boldsymbol{\beta}, \boldsymbol{z}, \boldsymbol{\omega}; \boldsymbol{\gamma} ) & = \sum_{g=1}^G\Omega_g(\boldsymbol \beta_g) + \hat Q_\tau(\boldsymbol z) + \boldsymbol{\gamma}_1^\top\left(\boldsymbol{X}_1\boldsymbol{\beta}_1+\boldsymbol{z}+\boldsymbol{\omega}_2+\cdots+\boldsymbol{\omega}_G-\boldsymbol{y}\right)                                                  \\
                                                                                                        & \hphantom{ = } +\frac{\phi}{2}\left\|\boldsymbol{X}_1 \boldsymbol{\beta}_1+\boldsymbol{z}+\boldsymbol{\omega}_2+\cdots+\boldsymbol{\omega}_G-\boldsymbol{y}\right\|_2^2 +\sum_{g=2}^G \boldsymbol{\gamma}_g^\top\left(\boldsymbol{X}_g \boldsymbol{\beta}_g-\boldsymbol{\omega}_g\right) \\
                                                                                                        & \hphantom{ = }+\frac{\phi}{2} \sum_{g=2}^G\left\|\boldsymbol{X}_g \boldsymbol{\beta}_g-\boldsymbol{\omega}_g\right\|_2^2,
    \end{aligned}
\end{equation}

where $\boldsymbol \gamma_g \in \mathbb R^n, g=1,\dots, G$ are dual vectors. The multi-block ADMM for \eqref{eq:fsadmm_lagrangian} consists of the update cycle $\boldsymbol \beta^{k} \to \boldsymbol \omega^{k+\frac{1}{2}} \to \boldsymbol z^{k} \to \boldsymbol \omega^{k+1} \to \boldsymbol \gamma^{k}$ given by

\begin{equation}
    \label{eq:fsadmm_update}
    \left\{\begin{array}{l}
        \boldsymbol{\beta}^{k+1}=\operatorname{argmin} \mathcal{L}_\phi\left(\boldsymbol{\beta}, \boldsymbol{z}^k, \boldsymbol{\omega}^k ; \boldsymbol{\gamma}^k\right)+\frac{\phi}{2}\left\|\boldsymbol{\beta}-\boldsymbol{\beta}^k\right\|_{\mathcal{T}_{\boldsymbol \beta}}^2   \\
        \boldsymbol{\omega}^{k+\frac{1}{2}}=\operatorname{argmin} \mathcal{L}_\phi\left(\boldsymbol{\beta}^{k+1}, \boldsymbol{z}^k, \boldsymbol{\omega} ; \boldsymbol{\gamma}^k\right)                                                                                             \\
        \boldsymbol{z}^{k+1}=\operatorname{argmin} \mathcal{L}_\phi\left(\boldsymbol{\beta}^{k+1}, \boldsymbol{z}, \boldsymbol{\omega}^{k+\frac{1}{2}} ; \boldsymbol{\gamma}^k\right)+\frac{\phi}{2}\left\|\boldsymbol{z}-\boldsymbol{z}^k\right\|_{\mathcal{T}_{\boldsymbol z}}^2 \\
        \boldsymbol{\omega}^{k+1}=\operatorname{argmin} \mathcal{L}_\phi\left(\boldsymbol{\beta}^{k+1}, \boldsymbol{z}^{k+1}, \boldsymbol{\omega} ; \boldsymbol{\gamma}^k\right)                                                                                                   \\
        \boldsymbol{\gamma}_1^{k+1}=\boldsymbol{\gamma}_1^k+\xi \phi\left(\boldsymbol{X}_1 \boldsymbol{\beta}_1^{k+1}+\boldsymbol{z}^{k+1}+\sum_{g=2}^G \boldsymbol{\omega}_g^{k+1}-\boldsymbol{y}\right)                                                                          \\
        \boldsymbol{\gamma}_g^{k+1}=\boldsymbol{\gamma}_g^k+\xi \phi\left(\boldsymbol{X}_g \boldsymbol{\beta}_g^{k+1}-\boldsymbol{\omega}_g^{k+1}\right), \quad g=2, \ldots, G
    \end{array}\right.
\end{equation}
where $\xi \in (0, (\sqrt{5} + 1) / 2)$, and $\mathcal T_{\boldsymbol\beta}$ and $\mathcal T_{\boldsymbol z}$ are positive semi-definite matrices which are crucial for convergence. Note that $\boldsymbol \omega_g$ are updated twice for improved convergence.

\subsection{Some remarks on FS-QRADMM}

\label{subsec:parallel_fsadmm}
As we can see from \eqref{eq:fsadmm_lagrangian} and \eqref{eq:fsadmm_update}, by introducing slack variables $ \boldsymbol{\omega}_g $ as local copies of $ \boldsymbol{X}_g \boldsymbol{\beta}_g $ and a block-diagonal $\mathcal T_{\boldsymbol\beta}$, the augmented Lagrangian becomes separable in the block variables $\boldsymbol{\beta}_g$. Consequently, FS-QRADMM updates $\boldsymbol{\beta}_g$ via $G$ independent subproblems that can be computed in parallel across blocks. Conditional on $\{\boldsymbol{\beta}_g\}_{g=1}^G $ and $\boldsymbol{z}$, the updates of $\{\boldsymbol{\omega}_g\}_{g=2}^G$ and their duals $\{\boldsymbol{\gamma}_g\}_{g=1}^G$ are likewise block-wise separable and amenable to parallel execution. By leveraging this parallelism, FS-QRADMM effectively alleviates the computational bottleneck associated with very large
$p$, achieving both high computational efficiency and accurate coefficient estimation (\cite{wenFeaturesplittingAlgorithmsUltrahigh2025}).

\medskip

However, relative to a two-block ADMM scheme, FS-QRADMM's update cycle creates a longer reliance chain with multiple stages, which entails extra updates of the slack variables $\boldsymbol \omega_g$ and duals $\boldsymbol \gamma_g$ for $g=2, \dots, G$ (where $\boldsymbol \gamma_1$ corresponds to the single dual vector in the two-block scheme). This incurs additional memory burden of maintaining $2(G-1)$ vectors of dimension $n$. Moreover, in a parallel setting, the effective computational speed at each stage is governed by the slowest of the $G$ blocks, together with the cost of aggregating and disseminating global quantities such as $\sum_{g=1}^G\boldsymbol X_g\boldsymbol\beta_g$ and $\boldsymbol z$ for synchronization. Therefore, even though the update of each $\boldsymbol \omega_g$ or $\boldsymbol \gamma_g$ is algebraically simple, the stage that updates all of them tends to lengthen as $G$ grows. As a result, the combined memory, computational and coordination costs can erode the gains from parallelizing the $\boldsymbol \beta$-updates when $G$ increases or when $n$ is very large. A natural question would be: is it possible to avoid the introduction of extra $\boldsymbol \omega_g$ and $\boldsymbol \gamma_g$ to further reduce the computational and memory costs? We will focus next on answering this question.

\medskip

\section{A Feature-Splitting Proximal Point Algorithm for PQR}
\label{sec:FS-QRPPA}
Recently, \cite{wuFeatureSplittingParallel2025} addressed a challenge analogous to that discussed in Section \ref{subsec:parallel_fsadmm}, specifically within the framework of feature-splitting Dantzig selectors solved via multi-block ADMM (\cite{wenNonconvexDantzigSelector2024}). To mitigate the computational burden, they developed a feature-splitting proximal point algorithm (PPA). Relative to multi-block ADMM, their PPA formulation substantially reduces the number of auxiliary variables and simplifies the update logic. PPA (\cite{martinetBreveCommunicationRegularisation1970}, \cite{rockafellarMonotoneOperatorsProximal1976a}) constitutes a broad family of iterative methods for solving optimization problems where optimality conditions are expressed as generalized equations. This framework encompasses nonsmooth convex optimization, rendering it directly applicable to our focus on QR with a weighted $\ell_1$ penalty. PPA underpins many statistical optimization methods. See \cite{polsonProximalAlgorithmsStatistics2015} for a comprehensive discussion. Notably, ADMM can be viewed as a special case of PPA (\cite{gabayApplicationsMethodMultipliers1983}, \cite{ecksteinDouglasRachfordSplitting1992}, Section 3.5 in \cite{boydDistributedOptimizationStatistical2010}).

\medskip

Motivated by the work of \cite{wuFeatureSplittingParallel2025}, we propose a feature-splitting PPA (FS-QRPPA) for PQR. Similar to the canonical two-block ADMM, our approach employs only $2n$ auxiliary variables, which may lead to improvements in both memory efficiency and computational cost. FS-QRPPA requires only the variables $\boldsymbol \beta \in \mathbb R^{p+1}$, $\boldsymbol z \in \mathbb R^n$, and $\boldsymbol \theta \in \mathbb R^n$ as in the canonical two-block ADMM formulation, and has a streamlined update scheme, with steps per-iteration mirroring the ADMM updates in \eqref{eq:admm_update}. It thereby avoids the additional memory and computational overhead encountered in FS-QRADMM, as discussed in Section~\ref{subsec:parallel_fsadmm}. Further details on PPA, and how FS-QRPPA fits into the sub-category called variable-metric PPA (VMPPA), are provided in Appendix D.

We now present the updating framework of FS-QRPPA, which can be summarized as follows:

\begin{equation}
    \label{eq:fsppa_update}
    \begin{cases}
        \begin{aligned}
            \boldsymbol \beta^{k+1}_g & =\argmin_{\boldsymbol \beta_g \in \mathbb R^{p_g}} \Omega_g(\boldsymbol \beta_g) - \boldsymbol \theta^{k}{}^\top \boldsymbol X_g  \boldsymbol\beta_g + \frac{1}{2}\| \boldsymbol\beta_g -  \boldsymbol\beta_g^k\|^2_{\boldsymbol M_g}, \quad g = 1, \dots, G              \\
            \boldsymbol z^{k+1}       & =\argmin_{\boldsymbol z \in \mathbb R^n} \hat Q_\tau(\boldsymbol z) - \boldsymbol\theta^k{}^\top \boldsymbol z + \frac{\mu}{2}\|\boldsymbol z - \boldsymbol z^k\|^2_2                                                                                                     \\
            \boldsymbol \theta^{k+1}
                                      & = \boldsymbol \theta^k - \dfrac{\mu}{G+1}\left[2\left(\sum_{g=1}^G \boldsymbol X_g \boldsymbol \beta^{k+1}_g + \boldsymbol z^{k+1} - \boldsymbol y\right) - \left(\sum_{g=1}^G \boldsymbol X_g \boldsymbol \beta^{k}_g + \boldsymbol z^{k} - \boldsymbol y\right)\right].
        \end{aligned}
    \end{cases}
\end{equation}

Here $\mu > 0$ is some pre-specified augmented parameter, $\boldsymbol M_g$ is some positive definite matrix such that $\boldsymbol M_g - \boldsymbol X_g^\top \boldsymbol X_g \succ 0$. Throughout this work, we take $\boldsymbol M_g = \eta_g\boldsymbol I_{p_g}$ where $\eta_g > \mu\Lambda_{\max}(\boldsymbol X_g^\top \boldsymbol X_g),\ g=1, \dots, G$. One may worry that it is costly to perform eigen-decomposition of $\boldsymbol X_g^\top \boldsymbol X_g$. However, highly-efficient algorithm like the power method (Chapter 8.2, \cite{golubMatrixComputations2013}) or the Lanczos iteration (Chapter 10.1, \cite{golubMatrixComputations2013}) can be deployed to get the top eigenvalue and mitigate the concern of excessive cost of a full eigendecomposition. Under the choice $\boldsymbol M_g = \eta_g\boldsymbol I_{p_g}$, the update for $\boldsymbol \beta^{k+1}$ can be further simplified to
\begin{equation}
    \label{eq:beta_argmin}
    \boldsymbol \beta^{k+1}_g =\argmin_{\boldsymbol \beta_g} \Omega_g(\boldsymbol \beta_g) - \boldsymbol \theta^{k}{}^\top \boldsymbol X_g  \boldsymbol\beta_g + \frac{\eta_g}{2}\| \boldsymbol\beta_g -  \boldsymbol\beta_g^k\|^2_2, \quad g=1, \dots, G.
\end{equation}

One favorable property of the update for  $\boldsymbol \beta^{k+1}$ in \eqref{eq:beta_argmin} and the update for $\boldsymbol z^{k+1}$ in \eqref{eq:fsppa_update} is that they both have an explicit form. Indeed, in the context of weighted $\ell_1$ penalty, for $g = 1, \dots, G$,
\begin{equation*}
    \begin{aligned}
        \boldsymbol \beta^{k+1}_g
        =\argmin_{\boldsymbol \beta_g} \|\boldsymbol \lambda_g\circ \boldsymbol \beta_g\|  - \boldsymbol \theta^{k}{}^\top \boldsymbol X_g  \boldsymbol\beta_g + \frac{\eta_g}{2}\| \boldsymbol\beta_g -  \boldsymbol\beta_g^k\|^2_2
    \end{aligned}
\end{equation*}
which implies
\begin{equation}
    \label{eq:beta_update}
    \boldsymbol \beta_g^{k+1} = \frac{1}{\eta_g}\left[\left(\eta_g \boldsymbol \beta_{g}^k + \boldsymbol X_g^\top \boldsymbol\theta^k -  \boldsymbol \lambda_{ g}\right)_+ -  \left(-\eta_g \boldsymbol \beta_{g}^k - \boldsymbol X_g^\top \boldsymbol\theta^k -  \boldsymbol \lambda_{ g}\right)_+\right]
\end{equation}
and
\begin{equation}
    \label{eq:z_update}
    \begin{aligned}
        \boldsymbol z^{k+1}
         & = \left( \boldsymbol z^k + \frac{1}{\mu}\boldsymbol\theta^k - \frac{\tau}{n\mu} \right)_+ - \left( - \boldsymbol z^k - \frac{1}{\mu}\boldsymbol\theta^k - \frac{1 - \tau}{n\mu} \right)_+.
    \end{aligned}
\end{equation}

As demonstrated in \eqref{eq:beta_update}, the structural essence of FS-QRPPA lies in its block-separability. Similar to FS-QRADMM, the update for each block $\boldsymbol \beta_g^{k+1}$ depends solely on local data $(\boldsymbol X_g, \boldsymbol \beta_g^{k})$. This decoupling enables parallel execution across $G$ independent processors or local machines. Similarly, the computation of $\boldsymbol X \boldsymbol \beta$, which is required for the update of $\boldsymbol\theta^{k+1}$, can be divided into $G$ simultaneous sub-tasks of calculating $\boldsymbol X_g\boldsymbol \beta_g$, followed by computing their sum. Such aggregation step is cheap as it only takes the sum of $G$ $n$-dimensional vectors. Beyond this parallel efficiency, feature-splitting also accelerates by improving conditioning for the $\boldsymbol \beta_g$-subproblem, a property we elaborate on in Remark \ref{remark:choice_G}. We summarize the update scheme of FS-QRPPA for weighted $\ell_1$ penalty in Algorithm~\ref{alg:FS-QRPPA}, which formally defines the routine $\operatorname{FS-QRPPA}(\cdot)$.

\begin{algorithm}[ht]
    \caption{FS-QRPPA with weighted $\ell_1$ penalty ($\operatorname{FS-QRPPA}(\boldsymbol X, \boldsymbol y, \boldsymbol \beta^0, \boldsymbol z^0, \boldsymbol \theta^0, \mu, \{\boldsymbol \lambda_g\}_{g=1}^G, \{\eta_g\}_{g=1}^G)$)}
    \label{alg:FS-QRPPA}
    \KwIn{$\boldsymbol X = (\boldsymbol X_1, \dots, \boldsymbol X_G)$, $\boldsymbol y$, $\boldsymbol\beta^0 = \left(\boldsymbol\beta^0_1{}^\top, \dots, \boldsymbol\beta^0_G{}^\top\right)^\top$, $\boldsymbol z^0, \boldsymbol \theta^0$, $\mu$, $\{\boldsymbol \lambda_g\}_{g=1}^G$, $\{\eta_g\}_{g=1}^G)$}
    $k \gets 0$ \;
    \While{Stopping criteria not satisfied\footnotemark}
    {
        \ParFor{$g = 1$ \KwTo $G$}{
            Update $\boldsymbol \beta^{k+1}_g$ according to \eqref{eq:beta_update}\;
        }
        Update $\boldsymbol z^{k+1}$ according to \eqref{eq:z_update}\;
        Update $\boldsymbol \theta^{k+1}$ according to \eqref{eq:fsppa_update}\;
        $k \gets k+1$\;
    }
    \KwOut{$\boldsymbol w^{k} = (\boldsymbol \beta^{k}, \boldsymbol z^{k}, \boldsymbol \theta^{k})$ where $\boldsymbol \beta^{k} = \left(\boldsymbol \beta_1^{k}{}^\top, \dots,\boldsymbol \beta_G^{k}{}^\top \right)^\top$}
\end{algorithm}
\footnotetext{The details on stopping criteria are given in Appendix D.}

\begin{remark}
    \label{remark:choice_G}
    The number of partitions $G$ imposes a trade-off on the convergence behavior of Algorithm \ref{alg:FS-QRPPA}, mirroring effects observed in \cite{wenFeaturesplittingAlgorithmsUltrahigh2025}. On the one hand, a larger value of $G$ implies a finer partition, which significantly reduces the local maximum eigenvalue $\Lambda_{\max}(\boldsymbol X_g^\top \boldsymbol X_g)$ relative to its global counterpart $\Lambda_{\max}(\boldsymbol X^\top \boldsymbol X)$ required by two-block proximal ADMM approaches (\cite{guADMMHighDimensionalSparse2018}). Consequently, $\eta_g$ can be much smaller than its non-feature-splitting alternative. This reduction directly accelerates the $\boldsymbol\beta$-subproblem by permitting significantly larger step sizes. On the other hand, for a given $\mu$, the step size for the $\boldsymbol \theta$-subproblem is inversely proportional to $G$, so its update is likely to be slower as $G$ increases. In addition, synchronization overhead in parallel execution tends to grow with $G$. Therefore, a moderate value of $G$ (e.g. 5-50) is preferable to balance the efficiency of the two subproblems, subject to computational resources. Further discussion on the two-fold impact of $G$ on the convergence rate is provided in Remark A4 in Appendix E.
\end{remark}

\begin{remark}
    \label{remark:matrix_mg}
    The choice of $\boldsymbol M_g = \eta_g \boldsymbol I_{p_g}$ is motivated by computational considerations. Specifically, we only need to compute $\Lambda_{\max}(\boldsymbol X_g^\top \boldsymbol X_g)$, which can also be calculated in parallel. However, this choice is not necessarily optimal. There often exists some diagonal matrix $\tilde{\boldsymbol M}_g$  providing a tighter bound, i.e., $\eta_g \boldsymbol I_{p_g}  \succ\tilde{\boldsymbol M}_g$ and $\tilde{\boldsymbol M}_g \succ \boldsymbol X_g^\top \boldsymbol X_g$, which could accelerate convergence. However, finding such tighter $\tilde{\boldsymbol M}_g $ in general is non-trivial and is in fact an important problem in the area of image processing called finding the ``diagonal majorizers" (\cite{muckleyFastParallelMR2015}, \cite{mcgaffinAlgorithmicDesignMajorizers2015}). We leave the better design of $\boldsymbol M_g$ for future work.
\end{remark}

\medskip

\section{Q-linear convergence for FS-QRPPA}
\label{sec:Q-linear}
Leveraging recent advances in VMPPA, we establish the convergence properties of Algorithm \ref{alg:FS-QRPPA}. Specifically, we prove that the algorithm is globally convergent and, notably, achieves a Q-linear convergence rate. For clarity, we briefly recall the definitions of convergence rates (see, e.g., \cite{ortegaIterativeSolutionNonlinear2000}, \cite{nocedalNumericalOptimization2006}). In our context, a sequence $\{\boldsymbol w^k\}$ is said to converge Q-linearly to a solution set $\mathcal S$ if the quotient of successive errors is asymptotically bounded below 1. By contrast, a sequence converges R-linearly if the error is dominated by a Q-linearly converging sequence; or, equivalently, if the error is bounded by a geometrically decaying sequence.

The convergence properties of FS-QRPPA are summarized in the following theorem with technical details deferred to Appendix E.

\begin{theorem}
    \label{theorem:main_convergence}
    Denote $\left\{\boldsymbol w^k = \left(\boldsymbol\beta^k, \boldsymbol z^k, \boldsymbol\theta^k\right)\right\}$ the sequence generated by Algorithm \ref{alg:FS-QRPPA}.
    \begin{enumerate}[label=\arabic*.]
        \item (Algorithm convergence) It converges to some $\bar{\boldsymbol w} =\left(\bar{\boldsymbol\beta}, \bar{\boldsymbol z}, \bar{\boldsymbol\theta} \right) \in \mathcal S$, the set of saddle points of Lagrangian \eqref{eq:main_lagrangian}.
        \item (Q-linear convergence rate in $\boldsymbol H$-induced distance) There exists some positive definite matrix $\boldsymbol H$ depending only on $\boldsymbol X,\mu,G,\{\eta_g\}_{g=1}^G$ and a constant $r\in(0,1)$ such that
              \begin{equation*}
                  \lim_{k\to \infty} \dist_{\boldsymbol H}(\boldsymbol w^{k}, \mathcal S) = 0 \ \mathrm{ with } \ \limsup_{k\rightarrow\infty} \frac{\dist_{\boldsymbol H}(\boldsymbol w^{k+1}, \mathcal S)}{\dist_{\boldsymbol H}(\boldsymbol w^{k}, \mathcal S)} \le r.
              \end{equation*}
    \end{enumerate}
\end{theorem}

\begin{corollary}[R-linear convergence rate in Euclidean distance]
    \label{coro:r_linear}
    Let $\{\boldsymbol w^{k}\}$ be the sequence in Theorem \ref{theorem:main_convergence}. There exists some $C > 0$ and $\rho \in (0, 1)$ such that for all $k = 1, 2, \dots$,
    \begin{equation*}
        \dist(\boldsymbol w^{k}, \mathcal S) \le C \rho^{k}.
    \end{equation*}
\end{corollary}

\begin{remark}
    The weighted $\ell_1$ penalty constitutes a special case of the weighted Elastic-Net penalty (\cite{zouAdaptiveElasticnetDiverging2009}, \cite{hoWeightedElasticNet2015}), recovered when the quadratic regularization parameters vanish. Algorithm \ref{alg:FS-QRPPA} admits a straightforward extension to accommodate PQR with the weighted Elastic-Net penalty. Under this generalization, Proposition \ref{prop:min_saddle_points} and Theorem \ref{theorem:main_convergence} remain valid. See Appendix B for details.
\end{remark}

\section{Extension to Non-Convex Penalty}
\label{sec:non-convex}
Algorithm \ref{alg:FS-QRPPA} can be seamlessly generalized to feature-splitting QR penalized by a general folded concave penalty $p_{\lambda}(|\beta|), \beta \in \mathbb R$ (\cite{fanStrongOracleOptimality2014}). General folded-concave penalties often yield more accurate coefficient estimates than the standard LASSO (\cite{brehenyCoordinateDescentAlgorithms2011}, \cite{fanStrongOracleOptimality2014}) as they avoid over-penalizing large coefficients. A further advantage of these penalties is that zero is not an ``absorbing state" (\cite{fanSureIndependenceScreening2008}): a feature suppressed to zero at an intermediate step may re-enter the active set in subsequent iterations. Prominent examples in this class include SCAD and MCP. The derivative $p_\lambda^\prime(|\beta|)$ takes specific forms depending on the choice of penalty. For SCAD with parameter $a > 2$, we have
\begin{equation}
    \label{eq:scad}
    p_\lambda^\prime(|\beta|) = \begin{cases}
        \lambda,                          & \text{if } |\beta| \le \lambda,          \\
        \frac{a\lambda - |\beta|}{a - 1}, & \text{if } \lambda < |\beta| < a\lambda, \\
        0,                                & \text{if } |\beta| \ge a\lambda,
    \end{cases}
\end{equation}
and for MCP penalty with parameter $a > 1$, we have
\begin{equation}
    \label{eq:mcp}
    p_\lambda^\prime(|\beta|) = \begin{cases}
        \lambda - \frac{|\beta|}{a}, & \text{if } |\beta| \le a\lambda, \\
        0,                           & \text{if } |\beta| > a\lambda.
    \end{cases}
\end{equation}

Correspondingly, we have $\Omega(\boldsymbol\beta) = \sum_{i=1}^p  p_\lambda(|\beta_i|)$, rendering the objective function \eqref{eq:vanilla_QR} nonconvex. While this implies the potential existence of multiple local minimizers, we can apply local linear approximation (LLA) (\cite{zouOnestepSparseEstimates2008}, \cite{wangQuantileRegressionAnalyzing2012}) to solve the optimization problem. LLA iteratively approximates the concave penalty with a local linear function, effectively transforming the non-convex problem into a sequence of weighted $\ell_1$ penalized sub-problems that are directly solvable via Algorithm \ref{alg:FS-QRPPA}. It has been shown that in a sparse QR setting and under mild conditions, the LLA estimator initialized at the LASSO converges to the oracle solution in only two iterations with overwhelming probability (see Corollary 8 in \cite{fanStrongOracleOptimality2014}). Consistent with this theory, \cite{wenFeaturesplittingAlgorithmsUltrahigh2025} report in numerical experiments that even a one-step LLA attains competitive accuracy. We summarize the $L$-step FS-QRPPA for SCAD or MCP penalties in Algorithm \ref{alg:FS-QRPPA-concave}, where all notational variants of $\boldsymbol\beta$, $\boldsymbol z$, and $\boldsymbol\theta$ follow the corresponding updates in \eqref{eq:beta_update}, \eqref{eq:z_update}, and \eqref{eq:fsppa_update}.

\begin{algorithm}[ht]
    \caption{FS-QRPPA with SCAD or MCP penalty}
    \label{alg:FS-QRPPA-concave}
    \KwIn{$\boldsymbol X = (\boldsymbol X_1, \dots, \boldsymbol X_G)$, $\boldsymbol y$, $\boldsymbol\beta^0 = (\boldsymbol\beta^0_1{}^\top, \dots, \boldsymbol\beta^0_G{}^\top)^\top$, $\boldsymbol z^0, \boldsymbol \theta^0$, $\lambda$, $\mu$, $\eta_g, g=1, \dots, G$}
    \For{$g = 1$ \KwTo $G$} {
        $\boldsymbol \lambda_g \gets \lambda \circ \boldsymbol 1$, where $\boldsymbol \lambda_g \in \mathbb R^{p_g}$;
    }
    $\tilde{\boldsymbol \beta}^{0}, \tilde{\boldsymbol z}^{0}, \tilde{\boldsymbol \theta}^{0} \gets \operatorname{FS-QRPPA}\left(\boldsymbol X, \boldsymbol y, \boldsymbol \beta^0, \boldsymbol z^0, \boldsymbol \theta^0, \mu, \{\boldsymbol \lambda_g\}_{g=1}^G, \{\eta_g\}_{g=1}^G\right)$ \;
    \For{$l = 1$ \KwTo $L$} {
    \For {$g = 1$ \KwTo $G$} {
    $\tilde{\boldsymbol \lambda}_g^{l} \gets \left(p_\lambda^\prime\left(|\tilde{\boldsymbol \beta}^{l-1}_{g, 1}|\right), \dots, p_\lambda^\prime\left(|\tilde{\boldsymbol \beta}^{l-1}_{g, p_g}|\right)\right)^\top$ with $p_\lambda^\prime(\cdot)$ being \eqref{eq:scad} for SCAD penalty or \eqref{eq:mcp} for MCP penalty\;
    }
    $\tilde{\boldsymbol \beta}^{l}, \tilde{\boldsymbol z}^{l}, \tilde{\boldsymbol \theta}^{l} \gets \operatorname{FS-QRPPA}(\boldsymbol X, \boldsymbol y, \tilde{\boldsymbol \beta}^{l-1}, \tilde{\boldsymbol z}^{l-1}, \tilde{\boldsymbol \theta}^{l-1}, \mu, \{\tilde{\boldsymbol \lambda}_g^{l}\}_{g=1}^G, \{\eta_g\}_{g=1}^G)$ \;
    }
    \KwOut{$\tilde{\boldsymbol \beta}^{L}$}
\end{algorithm}

\section{Numerical Studies}
\label{sec:numerical_studies}
To make the proposed FS-QRPPA framework accessible and practically usable for ultra-high-dimensional problems, we developed the open-source \texttt{R} package \texttt{fsQRPPA}. The package implements multi-core parallelization for both the LASSO (Algorithm \ref{alg:FS-QRPPA}) and non-convex (Algorithm \ref{alg:FS-QRPPA-concave}) penalized variants. This implementation fully leverages the parallel structure of FS-QRPPA, where computationally intensive operations, such as the matrix-vector products $\boldsymbol X\boldsymbol\beta$ and $\boldsymbol X^{\top}\boldsymbol\theta$, as well as the block-wise proximal updates for $\boldsymbol\beta$, are distributed across $G$ parallel processing blocks. In this section, we empirically validate the estimation and prediction performance as well as the scalability and computational efficiency of FS-QRPPA. Our evaluation begins with simulation studies under both sparse and dense settings. Furthermore, we apply \texttt{fsQRPPA} to ultra-high-dimensional genomic data from the UK Biobank. This application demonstrates the feasibility of conducting quantile regression at the biobank scale and highlights the utility of \texttt{fsQRPPA} in capturing heterogeneous associations. Throughout these numerical studies, we benchmark \texttt{fsQRPPA} against two state-of-the-art R packages for high-dimensional QR: \texttt{conquer} (\cite{tanHighDimensionalQuantileRegression2022}, \cite{heSmoothedQuantileRegression2023}, \cite{conquer}, \cite{manUnifiedAlgorithmPenalized2024}) and \texttt{rqPen} (\cite{rqPen}). Note that the FS-QRADMM algorithm discussed in Section \ref{sec:ADMM} is excluded from these comparisons as no public implementation is available.

\medskip

\texttt{conquer}, implemented using \texttt{RcppArmadillo}, is built upon convolution-type kernel smoothing of the pinball loss function, which enables the use of the gradient of the smoothed loss for acceleration. It employs a locally adaptive majorize–minimization scheme and has shown substantial speedups compared with other algorithms for QR such as the iterative coordinate descent method (\cite{pengIterativeCoordinateDescent2015}) and the two-block ADMM (\cite{guADMMHighDimensionalSparse2018}). By default, \texttt{conquer} uses a Gaussian kernel.

\medskip

\texttt{rqPen} wraps \texttt{hqreg} (\cite{yiSemismoothNewtonCoordinate2017}, \cite{hqreg}) as its backend solver for weighted $\ell_1$ penalized QR. Also powered by \texttt{RcppArmadillo}, \texttt{hqreg} smooths the pinball loss via Huber approximation and solves the resulting optimization problem using an efficient semi-smooth Newton coordinate-descent algorithm. It also applies an adaptive strong rule to screen inactive features for weighted $\ell_1$ penalized QR, which can remarkably accelerate computation in (ultra) high-dimensional settings along a penalty path $\{\lambda_t\}_{t=1}^T$. \texttt{rqPen} further extends \texttt{hqreg} to nonconvex penalties (SCAD and MCP) via LLA.

\medskip

For all algorithms, we supplied the standardized design matrix during model fitting, and transformed the estimated coefficients back to the original scale. Since both $\texttt{conquer}$ and $\texttt{rqPen}$ are hard-coded to use a one-step LLA for QR with concave penalties, we followed the same strategy for FS-QRPPA to ensure a fair comparison. In particular, we fixed $a = 3.7$ for SCAD and $a = 3$ for MCP.

\medskip

We employed a data-driven strategy to tune the hyperparameter $\lambda$, adapting our selection criterion to the sparsity structure of the problem. In sparse settings, we adopted the High-dimensional BIC criterion (HBIC) for QR proposed in \cite{pengIterativeCoordinateDescent2015}, namely:
\begin{equation}
    \operatorname{HBIC}(\lambda) = \log\left(\sum_{i=1}^n \rho_{\tau} (y_i - \boldsymbol x_i^\top \boldsymbol\beta)\right) + |\mathcal A| \frac{\log\log n}{n}C_n,
\end{equation}
where $|\mathcal A|$ is the cardinality of the active set, and $C_n > 0$ is a sequence of constants diverging to infinity as $n$ increases. As recommended by \cite{pengIterativeCoordinateDescent2015}, we set $C_n = \log p$.  Conversely, in dense settings characterized by numerous non-zero but weak signals, HBIC tends to select overly sparse and suboptimal models. To address this, we instead evaluated the predicted pinball loss on an independent validation set and selected the value of $\lambda$ that minimizes the validation loss. The sequence of candidate values of $\lambda$ was generated via the pivotal quantity approach proposed by \cite{belloniL1penalizedQuantileRegression2011} (see Appendix F for details). For FS-QRPPA, we implemented a warm-start strategy (\cite{friedmanPathwiseCoordinateOptimization2007}, \cite{friedmanRegularizationPathsGeneralized2010}), where the solution from the preceding $\lambda$ serves as the initialization for the current $\lambda$ (details in Appendix F).

\medskip

All simulations were conducted on a MacBook Pro with an M4 Max CPU and 64 GB of RAM. For the real-data analysis, computations were carried out on a high-performance computing cluster node with 48 CPU cores and 256 GB of RAM. In our numerical studies, we configured $G=5$ for simulations and $G=48$ for real-data analyses (reflecting the larger number of features in the real data applications).

\subsection{Simulations under a sparse setting}
\label{subsec:simulation_sparse}
We employed simulation settings similar to those in \cite{pengIterativeCoordinateDescent2015}, \cite{fanPenalizedQuantileRegression2021} and \cite{wenFeaturesplittingAlgorithmsUltrahigh2025} where the data come from an underlying heteroscedastic regression model:
\begin{equation}
    \label{eq:dgp_simulation}
    Y = X_{6} - X_{20} + X_{50} - X_{100} + X_{500} - X_{1000} + X_{p} + 0.7 X_{1}\epsilon,
\end{equation}
where $\epsilon \sim \mathcal N(0, 1)$. The design matrix $\boldsymbol X = [\boldsymbol 1, \boldsymbol X_1, \dots, \boldsymbol X_p]$ was generated using the following steps:
\begin{enumerate}
    \item Generate $\tilde{\boldsymbol X} = [\tilde{\boldsymbol X}_1, \dots,\tilde{\boldsymbol X}_p]$ from a multivariate normal distribution $\mathcal N(0, \boldsymbol \Sigma)$ where $\boldsymbol\Sigma \in \mathbb R^{p\times p}$ is an $\text{AR}(1)$ covariance matrix such that $\boldsymbol\Sigma_{ij} = 0.5^{|i - j|}$.
    \item Set $\boldsymbol X_1 = \Phi(\tilde{\boldsymbol X}_1)$ and $\boldsymbol X_j = \tilde{\boldsymbol X}_j, j = 2, \dots, p$, where $\Phi(\cdot)$ is the cumulative probability function of standard normal distribution.
\end{enumerate}

We considered three different quantiles, $\tau \in \{0.3, 0.5, 0.7\}$, under three distinct dimensional settings: $(n, p) = (1000, 2000)$, $(2000, 50000)$, and $(30000, 2000)$. For each scenario, we conducted 100 independent replications.
According to \eqref{eq:dgp_simulation}, the $\tau$-th conditional quantile of $Y$ is given by
\begin{equation}
    q_{\tau}(\boldsymbol x) =X_{6} - X_{20} + X_{50} - X_{100} + X_{500} - X_{1000} + X_{p} + 0.7X_{1}\Phi^{-1}(\tau).
\end{equation}
It follows that $X_1$ does not affect the conditional quantile of $Y$ when $\tau = 0.5$, but exerts a negative or positive influence when $\tau = 0.3$ or $0.7$, respectively.

\medskip

We evaluated the performance of feature selection and coefficient estimation accuracy of the aforementioned algorithms according to several metrics:
\begin{enumerate}
    \item Mean absolute error of coefficient estimation (AE): the average and standard deviation of the $\ell_1$ distance from the estimated coefficient vector to its true counterpart: $\sum_{j=1}^{p} |\hat\beta_j - \beta^*_j|$ over 100 replicates, where $\beta^*$ are the true coefficients. Note that $\beta_1^* = 0.7\Phi^{-1}(\tau)$.
    \item $P_1$: The proportion of fitted models where all active coefficients except $\beta_1$ are selected over 100 replicates. It is expected to be close to $1$ for all three selections of $\tau$.
    \item $P_2$: The proportion of fitted models where $\beta_1$ is selected over 100 replicates. It is expected to be close to 0 when $\tau = 0.5$ and close to 1 otherwise.
    \item Model size (Size): the average number of estimated active coefficients excluding the intercept over 100 replicates.
    \item Time: The running time (in seconds) of each algorithm along the path $\{\lambda_t\}_{t=1}^{50}$, averaged over 100 replicates.
\end{enumerate}

Simulation results are reported in Tables \ref{table:n1000p2000} - \ref{table:n30000p2000}, with the standard deviations of AE, Size, and Time shown in parentheses. We observe that, in terms of coefficient estimation accuracy and $P_2$, PQR with the MCP penalty consistently outperforms the LASSO and SCAD variants across all algorithms. Moreover, FS-QRPPA achieves the lowest AE among the competing methods for all three penalties. Regarding computational speed, FS-QRPPA exhibits performance comparable to \texttt{conquer} and \texttt{rqPen}, with potential improvements in the setting with $n = 2000$ and $p = 50000$, where the advantage of feature-splitting becomes more apparent. Notably, relative to a standard two-block ADMM without feature splitting (\cite{guADMMHighDimensionalSparse2018}), FS-QRPPA is substantially faster (see Appendix G).

\begin{table}[!htbp]
    \begin{threeparttable}
        \caption{Comparisons of algorithms for PQR when \( n=1000 \), \( p=2000 \); AE: the average and standard deviation of the $\ell_1$ distance from the estimated coefficient vector to its true counterpart; $P_2$: The proportion of fitted models where $\beta_1$ is selected over 100 replicates. It is expected to be close to 0; Size: the average number of estimated active coefficients excluding the intercept over 100 replicates; Time: The running time (in seconds) of each algorithm along the path $\{\lambda_t\}_{t=1}^{50}$, averaged over 100 replicates.\tnote{*}}
        \label{table:n1000p2000}
        \centering
        \begin{tabularx}{\textwidth}{l P{3em} Y P{3em} Y  Y }
            \toprule
            Algorithm         & \( \tau \) & AE            & P2   & Size        & Time        \\
            \midrule
            \textbf{FS-QRPPA} & 0.3        & 0.254 (0.083) & 0.98 & 8.42 (0.64) & 0.41 (0.08) \\
            \textbf{(LASSO)}  & 0.5        & 0.101 (0.034) & 0    & 7.18 (0.46) & 0.35 (0.01) \\
                              & 0.7        & 0.238 (0.088) & 0.98 & 8.45 (0.73) & 0.41 (0.06) \\
            \midrule
            \textbf{conquer}  & 0.3        & 0.523 (0.075) & 0.98 & 8.61 (0.71) & 0.20 (0.01) \\
            \textbf{(LASSO)}  & 0.5        & 0.265 (0.036) & 0    & 7.35 (0.58) & 0.17 (0.01) \\
                              & 0.7        & 0.498 (0.076) & 0.99 & 8.62 (0.79) & 0.20 (0.01) \\
            \midrule
            \textbf{rqPen}    & 0.3        & 0.525 (0.090) & 1    & 8.94 (1.08) & 0.25 (0.06) \\
            \textbf{(LASSO)}  & 0.5        & 0.259 (0.043) & 0    & 7.61 (0.76) & 0.23 (0.05) \\
                              & 0.7        & 0.492 (0.098) & 0.99 & 8.99 (0.96) & 0.24 (0.04) \\
            \midrule
            \textbf{FS-QRPPA} & 0.3        & 0.110 (0.042) & 1    & 8.00 (0.00) & 0.46 (0.04) \\
            \textbf{(SCAD)}   & 0.5        & 0.039 (0.012) & 0    & 7.00 (0.00) & 0.36 (0.02) \\
                              & 0.7        & 0.110 (0.047) & 1    & 8.00 (0.00) & 0.46 (0.05) \\
            \midrule
            \textbf{conquer}  & 0.3        & 0.231 (0.073) & 0.97 & 8.01 (0.27) & 0.23 (0.04) \\
            \textbf{(SCAD)}   & 0.5        & 0.052 (0.015) & 0    & 7.00 (0.00) & 0.38 (0.05) \\
                              & 0.7        & 0.232 (0.090) & 0.94 & 7.97 (0.30) & 0.23 (0.04) \\
            \midrule
            \textbf{rqPen}    & 0.3        & 0.188 (0.053) & 1    & 8.01 (0.10) & 1.21 (0.12) \\
            \textbf{(SCAD)}   & 0.5        & 0.037 (0.012) & 0    & 7.00 (0.00) & 1.15 (0.09) \\
                              & 0.7        & 0.181 (0.061) & 1    & 8.00 (0.00) & 1.24 (0.12) \\
            \midrule
            \textbf{FS-QRPPA} & 0.3        & 0.098 (0.054) & 1    & 8.01 (0.10) & 0.57 (0.06) \\
            \textbf{(MCP)}    & 0.5        & 0.039 (0.012) & 0    & 7.00 (0.00) & 0.42 (0.03) \\
                              & 0.7        & 0.108 (0.063) & 1    & 8.00 (0.00) & 0.56 (0.04) \\
            \midrule
            \textbf{conquer}  & 0.3        & 0.176 (0.066) & 0.99 & 8.01 (0.17) & 0.23 (0.05) \\
            \textbf{(MCP)}    & 0.5        & 0.052 (0.015) & 0    & 7.00 (0.00) & 0.38 (0.05) \\
                              & 0.7        & 0.174 (0.074) & 0.99 & 7.99 (0.10) & 0.23 (0.04) \\
            \midrule
            \textbf{rqPen}    & 0.3        & 0.157 (0.058) & 1    & 8.01 (0.10) & 1.27 (0.12) \\
            \textbf{(MCP)}    & 0.5        & 0.037 (0.012) & 0    & 7.00 (0.00) & 1.26 (0.09) \\
                              & 0.7        & 0.149 (0.064) & 1    & 8.00 (0.00) & 1.23 (0.09) \\
            \bottomrule
        \end{tabularx}
        \begin{tablenotes}[flushleft]
            \footnotesize
            \item[*] We do not report $P_1$ because all algorithms reach $P_{1} = 1$.
        \end{tablenotes}
    \end{threeparttable}
\end{table}

\begin{table}[!htbp]
    \centering
    \begin{threeparttable}
        \caption{Comparisons of algorithms for PQR when \( n=2000 \), \( p=50000 \); AE: the average and standard deviation of the $\ell_1$ distance from the estimated coefficient vector to its true counterpart; $P_2$: The proportion of fitted models where $\beta_1$ is selected over 100 replicates. It is expected to be close to 0; Size: the average number of estimated active coefficients excluding the intercept over 100 replicates; Time: The running time (in seconds) of each algorithm along the path $\{\lambda_t\}_{t=1}^{50}$, averaged over 100 replicates. \tnote{*}}
        \label{table:n20000p50000}
        \begin{tabularx}{\textwidth}{l P{3em} Y P{3em} Y  Y }
            \toprule
            Algorithms        & \( \tau \) & AE            & P2 & Size        & Time         \\
            \midrule
            \textbf{FS-QRPPA} & 0.3        & 0.219 (0.060) & 1  & 8.34 (0.50) & 14.61 (0.50) \\
            \textbf{(LASSO)}  & 0.5        & 0.105 (0.032) & 0  & 7.13 (0.37) & 12.25 (0.31) \\
                              & 0.7        & 0.218 (0.055) & 1  & 8.29 (0.50) & 14.92 (0.56) \\
            \midrule
            \textbf{conquer}  & 0.3        & 0.433 (0.048) & 1  & 8.28 (0.49) & 29.00 (0.69) \\
            \textbf{(LASSO)}  & 0.5        & 0.218 (0.023) & 0  & 7.14 (0.38) & 25.24 (0.58) \\
                              & 0.7        & 0.422 (0.049) & 1  & 8.38 (0.58) & 28.79 (0.70) \\
            \midrule
            \textbf{rqPen}    & 0.3        & 0.411 (0.058) & 1  & 8.50 (0.72) & 13.29 (1.09) \\
            \textbf{(LASSO)}  & 0.5        & 0.212 (0.031) & 0  & 7.16 (0.39) & 15.53 (1.37) \\
                              & 0.7        & 0.408 (0.054) & 1  & 8.38 (0.57) & 13.06 (1.07) \\
            \midrule
            \textbf{FS-QRPPA} & 0.3        & 0.075 (0.033) & 1  & 8.00 (0.00) & 15.05 (0.47) \\
            \textbf{(SCAD)}   & 0.5        & 0.029 (0.008) & 0  & 7.00 (0.00) & 12.98 (0.35) \\
                              & 0.7        & 0.074 (0.029) & 1  & 8.00 (0.00) & 15.23 (0.52) \\
            \midrule
            \textbf{conquer}  & 0.3        & 0.164 (0.045) & 1  & 8.02 (0.14) & 29.66 (0.86) \\
            \textbf{(SCAD)}   & 0.5        & 0.035 (0.010) & 0  & 7.00 (0.00) & 41.35 (2.54) \\
                              & 0.7        & 0.157 (0.049) & 1  & 8.02 (0.14) & 29.66 (0.99) \\
            \midrule
            \textbf{rqPen}    & 0.3        & 0.139 (0.037) & 1  & 8.01 (0.10) & 68.90 (2.98) \\
            \textbf{(SCAD)}   & 0.5        & 0.024 (0.008) & 0  & 7.00 (0.00) & 71.70 (2.16) \\
                              & 0.7        & 0.128 (0.039) & 1  & 8.00 (0.00) & 68.59 (2.74) \\
            \midrule
            \textbf{FS-QRPPA} & 0.3        & 0.055 (0.024) & 1  & 8.00 (0.00) & 16.83 (0.52) \\
            \textbf{(MCP)}    & 0.5        & 0.029 (0.008) & 0  & 7.00 (0.00) & 14.45 (0.36) \\
                              & 0.7        & 0.053 (0.022) & 1  & 8.00 (0.00) & 17.02 (0.53) \\
            \midrule
            \textbf{conquer}  & 0.3        & 0.121 (0.040) & 1  & 8.01 (0.10) & 29.74 (0.89) \\
            \textbf{(MCP)}    & 0.5        & 0.035 (0.010) & 0  & 7.00 (0.00) & 40.44 (2.78) \\
                              & 0.7        & 0.113 (0.042) & 1  & 8.00 (0.00) & 29.71 (0.90) \\
            \midrule
            \textbf{rqPen}    & 0.3        & 0.105 (0.035) & 1  & 8.01 (0.10) & 68.77 (2.86) \\
            \textbf{(MCP)}    & 0.5        & 0.024 (0.008) & 0  & 7.00 (0.00) & 73.19 (2.41) \\
                              & 0.7        & 0.095 (0.036) & 1  & 8.00 (0.00) & 68.31 (2.61) \\
            \bottomrule
        \end{tabularx}
        \begin{tablenotes}[flushleft]
            \footnotesize
            \item[*] We do not report $P_1$ because all algorithms reach $P_{1} = 1$.
        \end{tablenotes}
    \end{threeparttable}
\end{table}

\begin{table}[!htbp]
    \centering
    \begin{threeparttable}
        \caption{Comparisons of algorithms for PQR when \( n=30000 \), \( p=2000 \); AE: the average and standard deviation of the $\ell_1$ distance from the estimated coefficient vector to its true counterpart; $P_2$: The proportion of fitted models where $\beta_1$ is selected over 100 replicates. It is expected to be close to 0; Size: the average number of estimated active coefficients excluding the intercept over 100 replicates; Time: The running time (in seconds) of each algorithm along the path $\{\lambda_t\}_{t=1}^{50}$, averaged over 100 replicates. \tnote{*}}
        \label{table:n30000p2000}
        \begin{tabularx}{\textwidth}{l P{3em} Y P{3em} Y  Y }
            \toprule
            Algorithms        & \( \tau \) & AE            & P2   & Size        & Time         \\
            \midrule
            \textbf{FS-QRPPA} & 0.3        & 0.046 (0.011) & 1    & 8.20 (0.45) & 20.08 (0.65) \\
            \textbf{(LASSO)}  & 0.5        & 0.021 (0.004) & 0    & 7.08 (0.27) & 17.56 (0.69) \\
                              & 0.7        & 0.047 (0.011) & 1    & 8.22 (0.48) & 20.00 (0.81) \\
            \midrule
            \textbf{conquer}  & 0.3        & 0.084 (0.011) & 1    & 8.29 (0.59) & 7.50 (0.26)  \\
            \textbf{(LASSO)}  & 0.5        & 0.036 (0.004) & 0.01 & 7.16 (0.39) & 7.63 (0.18)  \\
                              & 0.7        & 0.084 (0.010) & 1    & 8.25 (0.48) & 7.38 (0.18)  \\
            \midrule
            \textbf{rqPen}    & 0.3        & 0.058 (0.011) & 1    & 8.23 (0.47) & 4.20 (0.08)  \\
            \textbf{(LASSO)}  & 0.5        & 0.029 (0.004) & 0    & 7.10 (0.30) & 4.15 (0.07)  \\
                              & 0.7        & 0.058 (0.011) & 1    & 8.22 (0.44) & 4.26 (0.10)  \\
            \midrule
            \textbf{FS-QRPPA} & 0.3        & 0.011 (0.005) & 1    & 8.00 (0.00) & 20.06 (0.59) \\
            \textbf{(SCAD)}   & 0.5        & 0.004 (0.001) & 0    & 7.00 (0.00) & 18.32 (0.66) \\
                              & 0.7        & 0.010 (0.005) & 1    & 8.00 (0.00) & 20.12 (0.76) \\
            \midrule
            \textbf{conquer}  & 0.3        & 0.033 (0.009) & 1    & 8.00 (0.00) & 15.56 (1.52) \\
            \textbf{(SCAD)}   & 0.5        & 0.008 (0.002) & 0    & 7.00 (0.00) & 19.29 (1.90) \\
                              & 0.7        & 0.033 (0.009) & 1    & 8.00 (0.00) & 15.83 (1.43) \\
            \midrule
            \textbf{rqPen}    & 0.3        & 0.013 (0.005) & 1    & 8.00 (0.00) & 26.92 (1.20) \\
            \textbf{(SCAD)}   & 0.5        & 0.009 (0.003) & 0    & 7.00 (0.00) & 26.69 (1.23) \\
                              & 0.7        & 0.012 (0.005) & 1    & 8.00 (0.00) & 27.43 (1.39) \\
            \midrule
            \textbf{FS-QRPPA} & 0.3        & 0.011 (0.005) & 1    & 8.00 (0.00) & 21.03 (0.85) \\
            \textbf{(MCP)}    & 0.5        & 0.004 (0.001) & 0    & 7.00 (0.00) & 18.86 (0.67) \\
                              & 0.7        & 0.010 (0.005) & 1    & 8.00 (0.00) & 21.21 (1.06) \\
            \midrule
            \textbf{conquer}  & 0.3        & 0.033 (0.009) & 1    & 8.00 (0.00) & 15.37 (1.37) \\
            \textbf{(MCP)}    & 0.5        & 0.008 (0.002) & 0    & 7.00 (0.00) & 18.81 (1.85) \\
                              & 0.7        & 0.033 (0.009) & 1    & 8.00 (0.00) & 15.55 (1.39) \\
            \midrule
            \textbf{rqPen}    & 0.3        & 0.013 (0.005) & 1    & 8.00 (0.00) & 26.43 (0.82) \\
            \textbf{(MCP)}    & 0.5        & 0.009 (0.003) & 0    & 7.00 (0.00) & 26.32 (1.13) \\
                              & 0.7        & 0.012 (0.005) & 1    & 8.00 (0.00) & 26.87 (1.07) \\
            \bottomrule
        \end{tabularx}
        \begin{tablenotes}[flushleft]
            \footnotesize
            \item[*] We do not report $P_1$ because all algorithms reach $P_{1} = 1$.
        \end{tablenotes}
    \end{threeparttable}
\end{table}

\medskip

Beyond feature selection and coefficient estimation accuracy, another important application of QR is to construct prediction intervals for the response variable. Accordingly, we evaluated the empirical performance of the prediction intervals produced by the different algorithms. For each of the three simulation settings, we generated an additional 100 test sets of size $0.5n \times p$ from the same data-generating process and utilized the 100 replicates described in the previous subsection as training sets. We then applied the same procedure used in our selection and estimation study:  fitting PQR with LASSO, SCAD, and MCP penalties at quantile levels $\tau = 0.1$ and $\tau = 0.9$ along a sequence of penalty values $\{\lambda_t\}_{t=1}^T$ constructed by the aforementioned pivotal approach, and selecting the optimal penalty parameter via HBIC.

\medskip

For each test observation $\boldsymbol x_i$, the predicted $\tau$-th quantile for $Y_i$ is
$\hat q_\tau(Y_i | \boldsymbol x_i)=\boldsymbol x_i^\top \hat{\boldsymbol\beta}(\tau)$,
where $\hat{\boldsymbol\beta}(\tau)$ is fitted on the training set with the penalty parameter selected by HBIC. For each algorithm, we construct the prediction interval $[\,\hat q_{0.1}(\boldsymbol x_i),\, \hat q_{0.9}(\boldsymbol x_i)\,]$. These intervals are assessed according to three metrics:
\begin{enumerate}
    \item Empirical coverage rate: the proportion of $y_i$ that fall within the interval. In particular, to enforce the non-crossing property, if for some $i$ we have $\hat q_{0.1}(\boldsymbol x_i) > \hat q_{0.9}(\boldsymbol x_i)$, we count it as a coverage failure. It is desirable that this rate be close to 0.80.
    \item Lower-tail miscoverage: the proportion of $y_i$ below $\hat q_{0.1}(\boldsymbol x_i)$. Ideally, it should be close to 0.10.
    \item Upper-tail miscoverage: the proportion of $y_i$ above $\hat q_{0.9}(\boldsymbol x_i)$. Ideally, it should be close to 0.10.
\end{enumerate}

Results under three $(n,p)$ settings are summarized in Figure~\ref{fig:coverage_summary_sparse}, with the corresponding test pinball losses reported in Tables A1 - A3 in Appendix G. Across settings, FS-QRPPA with the three penalties (LASSO, SCAD, MCP) attains a median empirical coverage close to the nominal level of 0.80, and the median lower- and upper-tail miscoverage rates are close to 0.10. In contrast, \texttt{conquer} and \texttt{rqPen} display a tendency toward over-coverage, particularly under non-convex penalties: the empirical coverage exceeds 0.80, while both tail rates are noticeably below 0.10. One potential explanation is that smoothing-induced bias is more pronounced at the extreme quantile levels (0.1 and 0.9) for these two approaches. As $n$ increases, all algorithms exhibit more stable coverage (smaller dispersion across replicates). Within each setting, the dispersion of the three metrics is broadly comparable across methods.
\begin{figure}[!htbp]
    \centering
    \includegraphics[width=\linewidth]{
        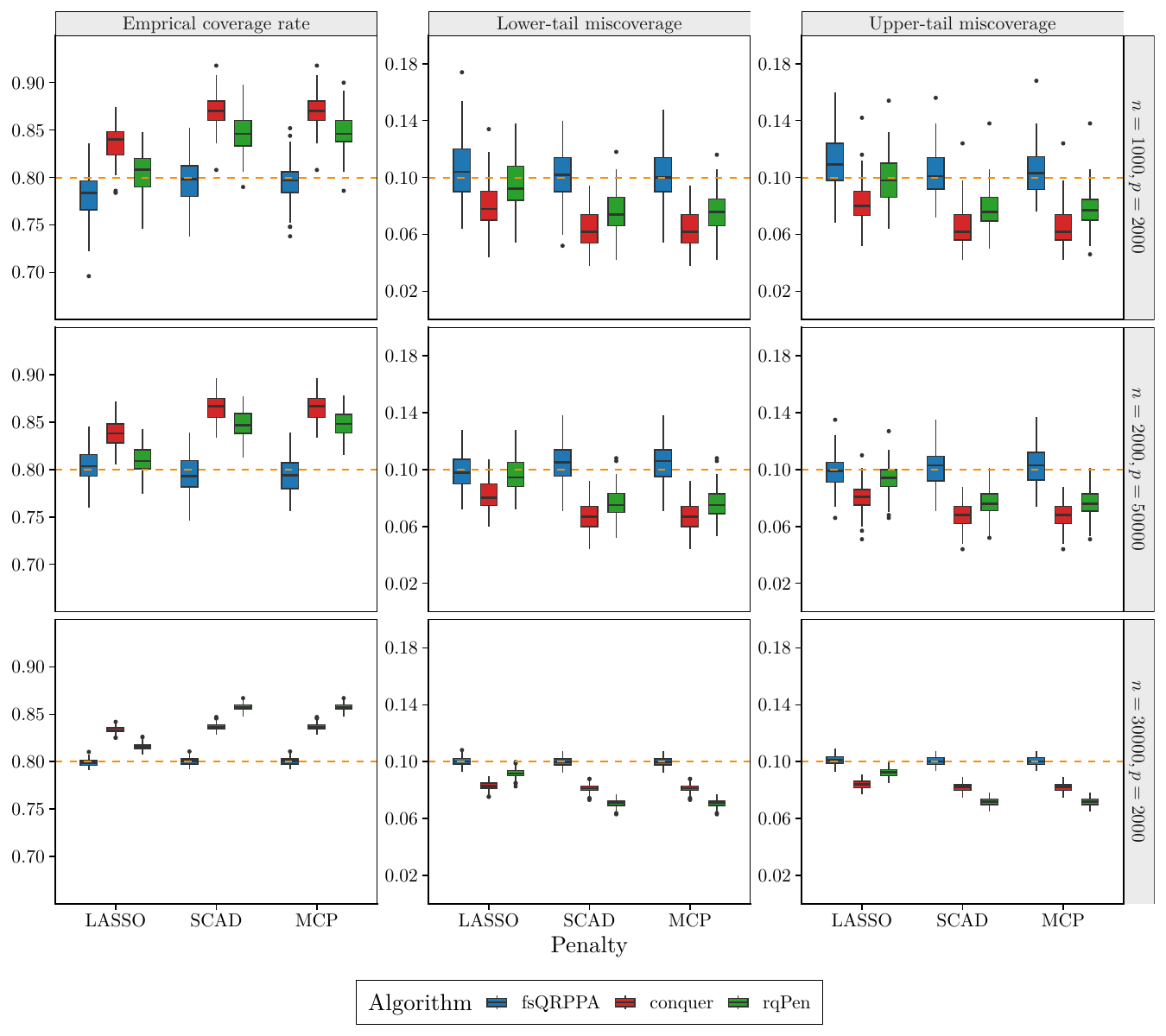}

    \caption{Comparison of empirical test-set coverage and tail proportions
        for predicted quantile bands ($\tau = 0.1$–$0.9$) with different penalties (LASSO, SCAD, MCP)
        under sparse settings with three different $(n,p)$. Empirical coverage rate: the proportion of $y_i$ that fall within the interval; Lower-tail miscoverage: the proportion of $y_i$ below $\hat q_{0.1}(\boldsymbol x_i)$; Upper-tail miscoverage: the proportion of $y_i$ above $\hat q_{0.9}(\boldsymbol x_i)$.}
    \label{fig:coverage_summary_sparse}
\end{figure}

\subsection{Simulations under a dense setting}
\label{subsec:simulation_dense}
We considered an additional simulation scenario under a dense setting, common in genetic studies where many complex traits (such as height) are highly polygenic, i.e. a large number of variants are weakly associated with the target phenotype. Since reliable variable selection and coefficient estimation are inherently ill-posed in such regimes, our primary objective is to evaluate the prediction accuracy at specific conditional quantiles.

To this end, we adapted the data-generating mechanism to a dense setting similar to that in \cite{wangRegenieQRSComputationallyEfficient2025}. Specifically, we consider the setting in which $n = 10000$ and $p = 20000$. The response is generated from a heteroscedastic regression model with random coefficients:
\begin{equation}
    Y = \sum_{j=1}^{m_1} a_j X_j + \left(\sum_{j=m_1 + 1}^{m_1 + m_2}|b_j| X_j\right)\epsilon,
\end{equation}
where $a_j \sim \mathcal N(0, \sigma_1^2/m_1)$ for $j = 1, \dots, m_1$, $b_j \sim \mathcal N(0, \sigma_2^2/m_2)$ for $j = m_1 + 1, \dots, m_1 + m_2$, and $\epsilon \sim \mathcal N(0, 1)$. We set $\sigma_1^2 = 0.3$ and $\sigma_2^2 = 0.1$, following \cite{wangRegenieQRSComputationallyEfficient2025}, and fix $m_1 = 500$ and $m_2 = 50$.  The design matrix was generated following the same procedure as in the sparse setting,  with the exception that in the second step, we set $\boldsymbol X_j = \Phi(\boldsymbol X_j)$ for $j = m_1 + 1, \dots, m_1 + m_2$.

\medskip

Here we focus on evaluating the empirical performance of the prediction intervals generated by the considered algorithms. The results over 100 independent replicates are summarized in Figure \ref{fig:coverage_summary_dense}, with the corresponding test pinball losses reported in Table A4 in Appendix G. Across all penalties, the three algorithms yield prediction intervals with slight under-coverage. This phenomenon is potentially due to regularization bias and limited capacity of high-dimensional QR models to accurately capture the many weak effects, and separate them from truly null effects. This limitation parallels findings in the polygenic risk score literature (\cite{zhaoPolygenicRiskScores2022}), which demonstrate that separating weak causal variants from null (noise) ones becomes increasingly intractable as the number of causal signals increases relative to the sample size.

\begin{figure}[!htbp]
    \centering
    \includegraphics[width=\linewidth]{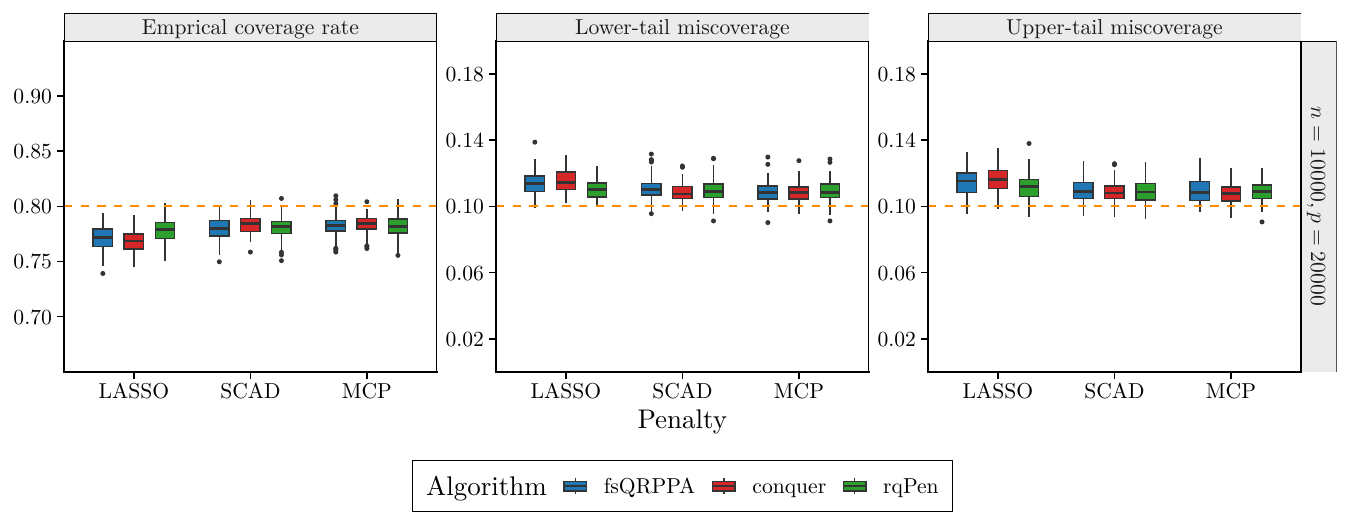}

    \caption{Comparison of empirical test-set coverage and tail proportions
        for predicted quantile bands ($\tau = 0.1$–$0.9$) with different penalties (LASSO, SCAD, MCP)
        in the dense setting. Empirical coverage rate: the proportion of $y_i$ that fall within the interval; Lower-tail miscoverage: the proportion of $y_i$ below $\hat q_{0.1}(\boldsymbol x_i)$; Upper-tail miscoverage: the proportion of $y_i$ above $\hat q_{0.9}(\boldsymbol x_i)$.}
    \label{fig:coverage_summary_dense}
\end{figure}

\subsection{Applications to UK Biobank data}
We return here to our motivating application on GWAS in UK Biobank. We focus on individuals of
European ancestry and two quantitative traits: height and lipoprotein(a). We analyzed data for 737401 genetic variants, a.k.a. Single Nucleotide Polymorphisms (SNPs), in 39699 individuals of European ancestry. Applying standard quality control procedures (\cite{RegenieQC}) to filter variants and samples, followed by removing linearly
dependent variants (known as Linkage Disequilibrium or LD pruning in genetics, s.t. $r^2 < 0.6$), yielded a final dataset with 482502 SNPs and 39699 individuals. Missing genotypes in this final dataset were imputed using mean-based imputation. In this (ultra) high-dimensional setting, both \texttt{conquer} and \texttt{rqPen} failed to complete on the cluster, terminating with errors for matrices of this size. Consequently, we report results exclusively for FS-QRPPA in this setting.

\medskip

There are no missing phenotypes in the height data. We split the data into training, validation, and test subsets with 30000, 5000, and 4699 individuals, respectively. For lipoprotein(a), the dataset comprises 30560 individuals, which we split into training, validation, and test subsets of size 25000, 3000, and 2560, respectively. Controlling for sex and age, we fitted FS-QRPPA on the respective training sets with LASSO, SCAD, and MCP penalties at quantile levels $\tau = 0.1$ and $\tau = 0.9$. As in simulations, we investigated the empirical coverage rate, the lower-tail, and upper-tail miscoverage rates. Additionally, we report the number of SNPs selected by the different models. These results are summarized in Table~\ref{table:real_data_height_lpa}, and the corresponding test pinball losses are reported in Tables~A5 and~A6 in Appendix~G.

\begin{table}[!htbp]
    \centering
    \caption{Applications to height and Lipoprotein(a). Comparison of empirical test-set coverage and tail proportions
        for predicted quantile bands ($\tau = 0.1$–$0.9$), and number of selected SNPs
        for FS-QRPPA prediction intervals with different penalties; Empirical coverage rate: the proportion of $y_i$ that fall within the interval; Below-0.1 rate: the proportion of $y_i$ below $\hat q_{0.1}(\boldsymbol x_i)$; Above-0.9 rate: the proportion of $y_i$ above $\hat q_{0.9}(\boldsymbol x_i)$.}
    \setlength{\tabcolsep}{5pt}
    \label{table:real_data_height_lpa}
    \begin{tabular}{l l ccccc}
        \toprule
        Trait
         & Penalty
         & \makecell[c]{Empirical                                          \\coverage rate}
         & \makecell[c]{Lower-tail                                         \\miscoverage}
         & \makecell[c]{Upper-tail                                         \\miscoverage}
         & \makecell[c]{\#SNPs                                             \\($\tau = 0.1$)}
         & \makecell[c]{\#SNPs                                             \\($\tau = 0.9$)} \\
        \midrule
        \multirow{3}{*}{Height}
         & LASSO                   & 0.769 & 0.119 & 0.112 & 21689 & 18213 \\
         & SCAD                    & 0.751 & 0.104 & 0.145 & 16086 & 16076 \\
         & MCP                     & 0.724 & 0.142 & 0.134 & 15076 & 12598 \\
        \midrule
        \multirow{3}{*}{Lipoprotein(a)}
         & LASSO                   & 0.793 & 0.095 & 0.112 & 16926 & 16692 \\
         & SCAD                    & 0.832 & 0.101 & 0.068 & 15077 & 8856  \\
         & MCP                     & 0.771 & 0.097 & 0.132 & 10479 & 10720 \\
        \bottomrule
    \end{tabular}
\end{table}

According to Table~\ref{table:real_data_height_lpa}, the empirical coverage rates of all models deviate modestly from the nominal level of $0.8$, indicating coverage error. For both traits, all models select on the order of ten thousand nonzero SNPs, suggesting that, in these applications to real data, prediction relies on a large set of small genetic effects rather than on a very sparse model, consistent with the known polygenic nature of these traits.

\paragraph*{FS-QRPPA reveals heterogeneous genetic effects across quantiles.}
A key advantage of QR is its capacity to reveal heterogeneous feature effects across the conditional distribution of a phenotype. To reliably detect such heterogeneity, we focused on genetic variants that remained active at both extremes of the distribution, specifically at quantile levels $\tau=0.1$ and $\tau=0.9$. We selected SNPs with nonzero estimated coefficients at both quantiles under FS-QRPPA across all penalties (LASSO, SCAD, and MCP). This screening yielded 226 candidate SNPs for height and 86 for lipoprotein(a).

\medskip

\begin{figure}[!htbp]
    \centering
    \includegraphics[width=\linewidth]{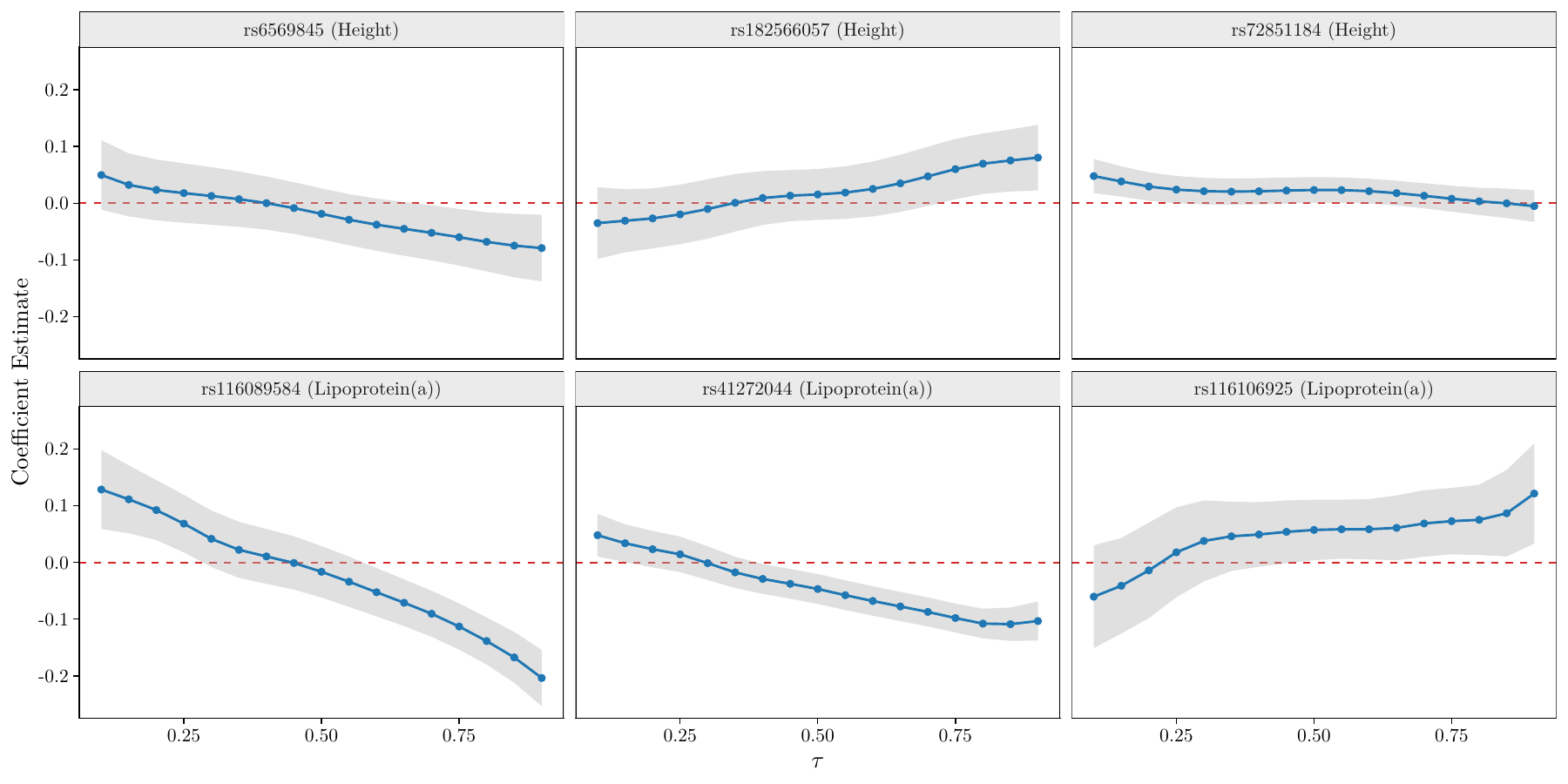}
    \caption{Heterogeneous effects of selected SNPs across quantile levels $\tau$ detected by FS-QRPPA for height (top row) and lipoprotein(a) (bottom row). The solid blue line connects the point estimates, the dashed red line marks zero, and the shaded region represents the 95\% asymptotic confidence interval.}
    \label{fig:hetero_height_lpa}
\end{figure}

Using these variants, we then fitted unpenalized QR on the training set via \texttt{conquer}, adjusting for sex and age, across a quantile grid from 0.1 to 0.9 (in increments of 0.05). For height, we identified 13 SNPs with opposite coefficient signs at the 0.1 and 0.9 quantiles, where at least one of the corresponding 95\% asymptotic confidence intervals excluded zero. Similarly, 14 SNPs exhibiting such sign differences were identified for lipoprotein(a). To illustrate this heterogeneity, Figure \ref{fig:hetero_height_lpa} displays the estimated coefficients and corresponding asymptotic confidence intervals across quantiles for three representative SNPs per trait. The plots for the remaining SNPs are provided in Appendix G.

\begin{figure}
    \centering
    \includegraphics[width=\linewidth]{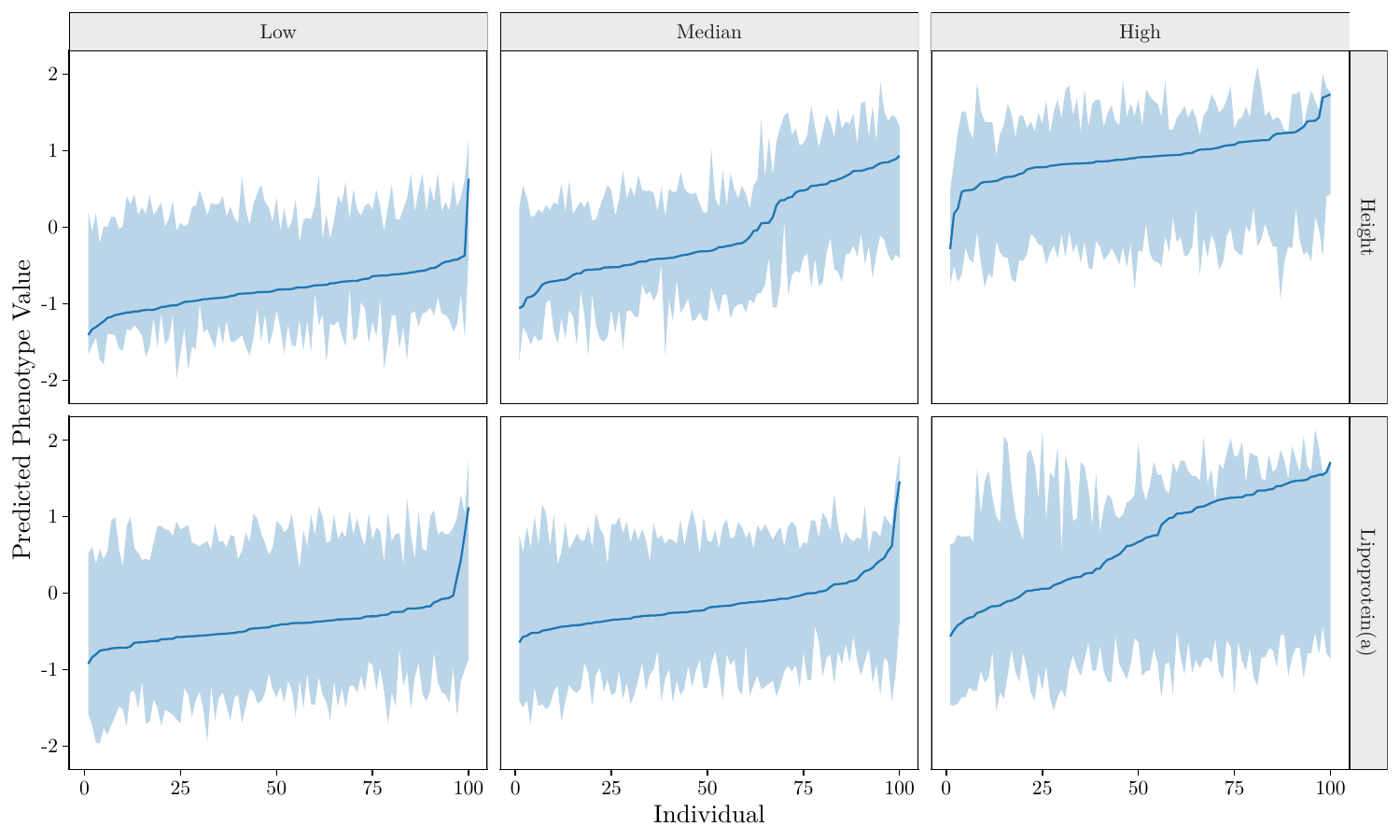}
    \caption{The 90\% prediction intervals and predicted median for height and lipoprotein(a) estimated using the LASSO penalty, stratified by phenotypic subgroup (Low, Median, and High). The solid blue line is the predicted median, and the light blue shaded region represents the 90\% prediction interval. Individuals within each panel are sorted by their predicted median.}
    \label{fig:pred_interval_lasso}
\end{figure}

\paragraph*{FS-QRPPA provides individualized prediction intervals.} Beyond identifying heterogeneous effects, another main advantage of QR is its ability to produce more realistic prediction intervals relative to conventional linear regression models. Specifically, we constructed prediction intervals for the held-out test data ($n=4699$ for height; $n=2560$ for lipoprotein(a)) using the coefficient vectors $\hat{\boldsymbol \beta}$ estimated on the training data across the LASSO, SCAD, and MCP penalties. We selected three representative subsamples of 100 individuals each: the low group (corresponding to the 100 smallest trait values), the median group (the 100 individuals closest to the median), and the high group (the 100 largest trait values). Figure \ref{fig:pred_interval_lasso} displays the LASSO-based prediction intervals for both traits. The results for SCAD and MCP are qualitatively similar and are provided in Appendix G. While standard linear models are often limited to symmetric prediction intervals of almost constant length across individuals, QR prediction intervals can be highly asymmetric and can vary in length from individual to individual, reflecting potential biological heterogeneity. Such individualized prediction intervals allow us to distinguish between people with narrow intervals, whose traits are mostly explained by genetic factors, and those with wider intervals, where non-genetic or environmental influences likely play a larger role.

\section{Discussion}
\label{sec:discussion}
In this work, we develop a feature-splitting proximal point algorithm (FS-QRPPA) for solving (ultra) high-dimensional penalized QR, with a theoretical linear convergence guarantee. We further provide a parallel implementation in the R package \texttt{fsQRPPA}, making penalized QR  tractable on datasets with on the order of billions of design-matrix entries. By exploiting feature-wise parallelism, FS-QRPPA alleviates memory and computational bottlenecks in modern (ultra) high-dimensional settings. This substantially broadens the applicability of QR to large-scale problems, such as genomic discovery and genomic trait prediction in GWAS.

\medskip

There are several meaningful directions for future research. Specifically, our current parallel implementation is optimized for a single compute node in a shared-memory setting. This design can already support applications to large GWAS analyses with tens of thousands of individuals. However, a single node may still be insufficient for full biobank-scale analyses (e.g. UK Biobank with $\sim 500$K individuals). It would therefore be practically important to develop a multi-node implementation of FS-QRPPA, enabling distributed-memory scaling and improving flexibility for modern large-scale QR. Complementary to this architectural enhancement, further acceleration is achievable by integrating safe screening techniques (\cite{ghaouiSafeFeatureElimination2011}, \cite{ndiaye2017gap}) which deterministically discard features whose regression coefficients $\beta$ are guaranteed to be zero (i.e. inactive features). Intuitively, such screening can substantially reduce the computational cost, particularly in sparse settings, and it is a natural direction for future work to incorporate these techniques into the FS-QRPPA scheme and to provide rigorous theoretical justification for the resulting improvements in computational efficiency. Beyond reducing the number of features, the convergence speed of FS-QRPPA could be considerably improved by optimizing the block structure. For instance, strategically permuting the columns of the design matrix to minimize the maximum eigenvalue of each block could effectively expedite the iterative updates. Taken together, these strategies form a synergistic roadmap for further improvements of FS-QRPPA.

\medskip

Additionally, our current feature-splitting framework is confined to PQR with the weighted Elastic-Net penalty. Specifically, our theoretical guarantee of linear convergence hinges on the piecewise linear-quadratic (PLQ) structure of this penalty. A worthwhile extension is the integration of general group-structured penalties, such as the Group LASSO (\cite{yuanModelSelectionEstimation2006}) or Sparse Group LASSO (\cite{simonSparseGroupLasso2013}). Such extensions are particularly relevant for genomic applications, where SNPs exhibit inherent grouping within genes. Indeed, the utility of group-penalized QR in genetic settings has been demonstrated in recent statistical literature (\cite{mendez-civietaAdaptiveSparseGroup2021}, \cite{ ouhouraneGroupPenalizedQuantile2022}). However, while the group-separability of these penalties preserves the computational feasibility of the feature-splitting architecture, they lack the PLQ property. Consequently, establishing convergence rates for these cases would necessitate a new theoretical analysis.

\medskip

Finally, in our simulation studies and real-data analyses we observed that QR prediction intervals can exhibit coverage error, with the realized coverage rate deviating from the nominal level. This phenomenon is not incidental; the prevalence of coverage bias has been rigorously investigated in the statistical literature (\cite{romanoConformalizedQuantileRegression2019, baiUnderstandingUndercoverageBias2021, gibbsCorrectingCoverageBias2025}), and recent work has highlighted the utility of conformal methods for genetic trait prediction (\cite{wangIndividualizedUncertaintyQuantification2025}). Importantly, the optimization scheme of FS-QRPPA is compatible with these correction strategies. A promising path for future work is to integrate conformal prediction into the FS-QRPPA pipeline and apply this unified framework to large-scale genetic data. This approach aims to provide accurate personalized predictions and uncertainty quantification, providing potential benefits in clinical applications.

\paragraph{Acknowledgments.} We thank Ma{\l}gorzata Bogdan for useful discussions on this topic. This research has been conducted using the UK Biobank Resource under Application Number 41849.

\paragraph{Funding.} We acknowledge support for this work from the Swedish Research Council.
\paragraph{Code Availability.} The \texttt{R} package \texttt{fsQRPPA} is available at \url{https://anonymous.4open.science/r/fsQRPPA-6764}, where FS-QRPPA with LASSO, SCAD and MCP penalties are supported.

\bibliographystyle{imsart-nameyear} 
\bibliography{references}       

\end{document}